\documentclass[onecolumn]{emulateapj}

\newcommand{\G}{\Gamma}

\newcommand{\sT}{\sigma_{\rm T}}

\newcommand{\e}{\epsilon}
\newcommand{\g}{\gamma}
\newcommand{\gp}{\gamma^{\,\prime}}

\newcommand{\kB} {k_{\rm B}}
\newcommand{\bcm}{\beta_{cm}}
\newcommand{\gcm}{\g_{cm}}

\newcommand{\Op}{\Omega^\prime}

\newcommand{\tgg}{\tau_{\g\g}}
\newcommand{\tp}{t^\prime}
\newcommand{\ep}{\epsilon^\prime}

\newcommand{\dD}{\delta_{\rm D}}

\newcommand{\psim}{\lower.5ex\hbox{$\; \buildrel \propto \over\sim \;$}}
\newcommand{\lbar}{\lower.0ex\hbox{$\; \buildrel {\lower0.0ex \hbox{-}} \over\lambda  \;$}}

\shorttitle{$\gamma$-ray FSRQ Blazar Analysis}
\shortauthors{Dermer, Finke, Krug, and B\"ottcher}

\begin{document}
\title {Gamma-Ray Studies of Blazars: Synchro-Compton Analysis of 
Flat Spectrum Radio Quasars}

\author{Charles D.\ Dermer,$^1$ Justin D. Finke,$^{1,2}$ Hannah Krug,$^{1,3}$ \& Markus B\"ottcher$^4$}
\affil{$^1$U.S.\ Naval Research Laboratory, Code 7653, 4555 Overlook SW, Washington, DC
  20375-5352\\
$^2$Naval Research Laboratory/National Research Council Postdoctoral Associate\\
$^3$Department of Astronomy, University of Maryland, College Park, MD 20742 \\
$^4$Astrophysical Institute, Department of Physics and Astronomy, 
	Ohio University, Athens, Ohio 45701
} 
\email{charles.dermer@nrl.navy.mil}

\begin{abstract}

We extend a method for modeling synchrotron and synchrotron 
self-Compton radiations
in blazar jets to include external Compton processes.  
The basic model assumption is that the blazar
radio through soft X-ray flux is nonthermal synchrotron radiation
emitted by isotropically-distributed electrons in the randomly
directed magnetic field of outflowing relativistic blazar jet plasma.
Thus the electron distribution is given by the synchrotron spectrum, depending only
on the Doppler factor $\delta_{\rm D}$ and mean magnetic field $B$,
given that the comoving emission
region size scale $R_b^\prime \lesssim c \dD t_v/(1+z)$, where $t_v$ is 
variability time and  $z$ is
source redshift. Generalizing the approach of Georganopoulos, Kirk,
and Mastichiadis (2001) to arbitrary anisotropic target radiation
fields, we use the electron spectrum implied by the synchrotron
component to derive accurate Compton-scattered $\gamma$-ray spectra
throughout the Thomson and Klein-Nishina regimes for external Compton
scattering processes.  We derive and calculate
accurate $\gamma$-ray spectra produced by
relativistic electrons that Compton-scatter (i) a point source of
radiation located radially behind the jet, (ii) photons from 
a thermal Shakura-Sunyaev accretion disk
and (iii) target photons from the central source
scattered by a spherically-symmetric shell of broad
line region (BLR) gas. Calculations of broadband spectral 
energy distributions from the radio through $\gamma$-ray regimes are presented,
which include self-consistent $\gamma\gamma$ absorption on the same 
radiation fields that provide target photons for Compton scattering. 
Application of this baseline flat spectrum
radio/$\gamma$-ray quasar model is considered in view of 
data from $\gamma$-ray telescopes and
contemporaneous multi-wavelength campaigns.

\end{abstract}


\section{Introduction}

A class of radio-loud blazar active galactic nuclei (AGNs) 
that emit luminous fluxes of $\gtrsim
100$ MeV -- GeV $\gamma$ rays was discovered with the Energetic
Gamma Ray Experiment Telescope (EGRET) on the {\em
Compton Observatory} \citep{har92,fic94,har99}. This result clarified
the nature of 3C 273, which was first identified as a $\gamma$-ray
emitting AGN in COS-B satellite data \citep{her77}.  The $\gamma$ rays
from blazars are certainly nonthermal in origin and associated
with the radio jets formed by the supermassive black holes that power
these sources. The largest subclass of EGRET AGNs are moderate redshift ($z\approx
1$) flat-spectrum radio quasars (FSRQs) with blazar properties, 
including apparent superluminal motion, rapidly variable optical
emission, high polarization, and intense broadened optical emission
lines.  Another subclass of the EGRET AGNs consists of $\gamma$-ray
emitting BL Lacertae (BL Lac) objects, which are generally at lower
redshifts ($z \approx 0.1$ -- $0.3$) and, by definition, have weak or absent
optical emission lines in their spectra. The X-ray selected BL Lac
(XBL) subset were discovered to be a class of TeV $\gamma$-ray sources
by the Whipple Observatory \citep{pun92,wee03}.

The common superluminal nature of the first identified $\gamma$-ray
blazars, namely 3C 273, 3C 279, and PKS 0528+134, led \citet{dsm92} to
propose a Compton-scattering origin for the $\gamma$ rays. In this
model, jet electrons Compton-scatter accretion-disk photons that
intercept the jet plasma. The nonthermal jet electrons can also
scatter internal synchrotron photons to produce a synchrotron
self-Compton (SSC) component \citep{bm96}. Given the broadened
emission lines in the spectra of FSRQs, accretion-disk radiation
scattered by surrounding gas of the broad line region (BLR) will
provide a further source of target photons to be scattered to
$\gamma$-ray energies \citep{sbr94}, as will radiation from a
surrounding dusty torus \citep{kat99,bsmm00}.  The accretion-disk and
scattered radiation will attenuate jet $\gamma$-rays through
$\gamma\gamma$ pair-production attenuation \citep{bk95,bl95}.

Expressions for the $\gamma$-ray spectral energy distributions (SEDs)
of blazars produced by Compton scattering processes have been derived
and calculated for many specific models of the black-hole/blazar jet
environment. In the case of external accretion-disk photons as the
target photon source, where the accretion disk is described by an
optically thick, geometrically thin thermal \cite{ss73} accretion
disk, Compton-scattered $\gamma$-ray spectra were calculated in the
Thomson regime by \citet{ds93,ds02}. Calculations of the
Thomson-scattered spectra for a quasi-isotropic target radiation field
formed by BLR gas or hot dust were made by \citet{sbr94}, \cite{dss97}
and \citet{bsmm00}.  Detailed numerical calculations including both
accretion disk and scattered radiation fields, have been made by, e.g.,
\citet{kt05,bb00} and \citet{br04}.

Compton scattering in the Klein-Nishina regime is not so simple to
treat compared to analyses restricted to the Thomson regime, but is
unavoidable for blazar analysis in the era of the {\em Fermi
Gamma-Ray Space Telescope} (formerly known as {\em GLAST}) and the
ground-based $\gamma$ Cherenkov telescopes. For surrounding isotropic
radiation fields in the stationary frame of the blazar AGN,
\citet{gkm01} suggested to transform the comoving electron
distribution to the stationary frame and then scatter the target
photons to $\gamma$-ray energies, using the formula first derived by
Jones \citep{jon68,bg70}.  This approach is generalized in this paper
to surrounding anisotropic radiation fields.

The usual spectral modeling approach proceeds by injecting power-law
electrons and evolving these particles while they produce the output
synchrotron and Compton-scattered radiation \citep[e.g.,][]{ds93,bms97}.  
For example, \citet{mod05} 
calculate electron energy evolution and spectral formation 
throughout the Thomson and Klein-Nishina regimes for different
ratios of synchrotron and isotropic radiation field energy 
densities. They show that reduced Compton losses 
in the Klein-Nishina regime compared to synchrotron losses 
can lead to spectral hardening of the synchrotron component
in the optical/X-ray regime \citep[noted earlier by][]{da02}.
%, as originally proposed to explain 
%spectral hardenings seen in X-ray knots of radio galaxies with Chandra
A difficulty in this
approach is that the electron energy-loss rate depends on the photon
spectrum of the comoving radiation field, not just the total radiation
energy density, and this field evolves with time. 
The modeler is faced with the prospect of
simultaneously fitting the synchrotron and Compton components.  The
acceleration scenario may well be over-simplified, and non-power-law
particle injection distributions could be more realistic than
power-law injection spectra, e.g., due to nonlinear effects in Fermi
acceleration. Moreover, a separation between the acceleration and
radiation zones may not be justified.

Here we extend a method of blazar analysis recently proposed for TeV
blazars \citep{fdb08} that avoids these difficulties.  For a standard
$\gamma$-ray blazar model, where isotropically distributed electrons
spiral in a randomly oriented magnetic field with mean magnetic field
strength $B$ in the fluid frame, the measured synchrotron flux
directly reveals the electron spectrum responsible for the synchrotron
radiation. The only uncertainties are the mean magnetic field $B$, the
comoving size scale $R_b^\prime$ of the emitting region, and the
Doppler factor $\dD = [\Gamma (1-\beta\cos\theta)]^{-1}$ ($\G =
1/\sqrt{1-\beta^2}$ is the bulk Lorentz factor of the outflow).  With
this electron spectrum, we then Compton-scatter target photons of the
surrounding radiation fields using the head-on approximation to the
total Compton cross section \citep{ds93}, valid when the electron
Lorentz factor $\gamma \gg 1$. This generalizes the approach of
\citet{gkm01} to surrounding anisotropic radiation fields.  The
temporally-evolving electron spectrum in blazars can be derived in
this approach from simultaneous multiwavelength blazar data. Values of
$B$, $\dD$, and jet power can then be deduced.  The related treatment
for XBLs applied to PKS 2155-304, including more details about the
derivation of the electron spectrum from the synchrotron component,
the derivation and calculations of the SSC component and internal
$\gamma$-ray opacity by the synchrotron photons, is given by
\citet{fdb08}.

Analysis of blazar SEDs using this approach is presented in Section 2,
where formulas to calculate Compton-scattered internal and external
radiation and a $\delta$-function approximation for $\gamma\gamma$
opacity from the internal radiation field are given. Derivations of
the Compton-scattered spectrum for specific examples of external
radiation fields consisting of a monochromatic point source of
radiation radially behind the jet, a Shakura-Sunyaev disk model, and a
model BLR radiation field are derived in Section
3.  Discussion of the results is found in Section 4.
%%%%%%%%%%%%%%%%%%%%%%%%%%%%%%%%%%%%%%%%%%%%%%%%%%%%%%%%%%%%%
%%%%%%%%%%%%%%%%%%%%%%%%%%%%%%%%%%%%%%%%%%%%%%%%%%%%%%%%%%%%

\section{One-Zone Synchrotron/Synchrotron Self-Compton Model with 
$\gamma\gamma$ Opacity}

We consider a one-zone model for blazar flares. Multiple zones could still be
allowed, but the product of the duty cycle and number of  
zones would have to be small enough that interference of 
emissions from the different regions would still permit rapid variability. 
In this case, the emission would still predominantly arise from a
single zone. Distinct zones could also emit the bulk of 
their radiation in different wavebands. In this case, the cospatiality assumption
often made in blazar modeling would not apply. 
In this regard, correlated variability data is essential to 
test the underlying assumptions made when a one-zone model 
is employed. Slowly varying radio/IR
synchrotron and hard X-ray and low-energy $\gamma$-ray Compton
emissions could involve extended emission regions.

A radiative event from the source emission region that varies on a
comoving timescale $\tp_v\gtrsim R^\prime_b/c$ is related to the
observed variability timescale through the relation $\tp_v = \dD
t_v/(1+z)$, where $z$ is redshift; thus the comoving blob radius is 
$R^\prime_b\lesssim c\dD t_v/(1+z)$. The inequality allows us to
neglect light-travel time effects from different parts of the emitting
volume and avoid integrations over source volume. Within this zone,
the nonthermal electrons with isotropic pitch angle distribution
 are described by the total comoving electron
number spectrum $N_e^\prime(\gp )$ in terms of comoving electron
Lorentz factor $\gp $. The
magnetic field is assumed to be randomly oriented in the comoving
fluid frame. The relativistic electrons that gyrate in this field
radiate nonthermal synchrotron radiation, observed as the low-energy
component in blazar SEDs.

\subsection{Synchrotron and Self-Compton Components}

The  $\nu F_\nu$ synchrotron radiation spectrum can
be approximated by the expression
\begin{equation}
f_\e^{syn} \cong {\dD^4\over 6\pi d_L^2} \; c\sigma_{\rm T} U_B 
\gamma_s^{\prime 3} N_e^\prime (\gp_s)\;,\;
\label{fes}
\end{equation}
where 
\begin{equation}
\gp_s =  \sqrt{{\e(1+z)\over \dD\epsilon_B}}=  \sqrt{{\ep \over \epsilon_B}}\;,
\label{gps}
\end{equation}
$d_L=d_L(z)$ is the luminosity distance, $c$ is the speed of
light, $\sigma_T$ is the Thomson cross-section, $z$ is the source
redshift, and the comoving magnetic-field energy density of the
randomly-oriented comoving field with comoving mean intensity
$B$ is $U_B \equiv B^2/8\pi\;.$ We use $\epsilon$ and
$\epsilon^\prime$ to refer to the dimensionless photon energy in the
observer and comoving frame, respectively.  Here and throughout this
paper, unprimed quantities refer to the observer's frame, and primed
quantities refer to the frame comoving with the AGN's jet, with the
exception being $B$, the comoving magnetic field.  Inverting this
expression gives the comoving electron distribution
\begin{equation}
N_e^\prime(\gp_s ) = V_b^\prime n_e(\gp_s ) 
\cong {6\pi d_L^2 f_{\e_{syn}}^{syn}\over c\sigma_{\rm T} U_B \dD^4\gamma_s^{\prime 3}}
\;,\;\;
\label{Neprimegps}
\end{equation}
where 
\begin{equation}
\e_{syn} = {\dD\e_B\g_s^{\prime 2}\over 1+z}\;,
\end{equation}
$\e_B = B/B_{cr}$ is the ratio of $B$ and the critical magnetic field
$B_{cr} = m_e^2 c^3/e\hbar \cong 4.41\times 10^{13}$ G \citep{ds02},
and $V_b^\prime=4\pi R^{\prime 3}_b/3$ is the comoving
volume of the blob.  Note that $U_B = \e_B^2 U_{B_{cr}} =\e_B^2
B_{cr}^2/8\pi$. Eq.  (\ref{Neprimegps}) gives a good representation to
the source electron distribution when the $\nu F_\nu$ spectral index
$a < 4/3$ (i.e., for spectra softer than $a = 4/3$, adopting the
convention $f_\e\propto \e^a$) and away from the high-energy
exponential cutoff of the spectrum \citep[see][for comparison]{fdb08}.

%The $\delta$-function approximation for the synchrotron
%photon emissivity $\dot n_{syn}(\e )$ (ph cm$^{-3}$ s$^{-1}$
%$\e^{-1}$) is given by
%\begin{equation}
%\dot n^{\prime }_{syn}(\ep) \cong {2\over 3}\; { c \sigma_{\rm T} u_B }
%\; \e^{\prime -1/2} \epsilon_B^{-3/2}\; n^\prime_e\big(\gp_s\big)\;\cong\;
%{2c\sigma_{\rm T}U_{B_{cr}}\over 3V_b^\prime m_ec^2}\;{N^\prime_e(\gp_s )
%\over  \gp_s}\;,
%\label{dotnsyne}
%\end{equation}
%where $\gp_s \equiv \sqrt{\ep/ \e_B}$ and $u_B \equiv U_B/m_ec^2$.
%The target synchrotron radiation field is therefore given by
%\begin{equation}
%n^\prime_{syn}(\ep ) = {u^\prime(\ep)\over m_ec^2 \ep}\cong {R_b^\prime\over c} \; \dot n_{syn}^\prime( \ep )\;\cong
% {3d_L^2 f_\e^{syn}\over 
%m_ec^3R^{\prime 2}_b \dD^4 \e_B^2 \g^{\prime 4}_s}\;.
%\label{nprimesynep}
%\end{equation}

The SSC $\nu F_\nu$ flux is given by
\begin{equation}
f_\e^{SSC} \; = \;{\dD^4\over d_L^2}\; \ep_s L^\prime_{SSC}(\ep_s,\Op_s )\;.
\label{feSSC}
\end{equation}
The formula of \citet{jon68} (see also \citet{bg70}) gives the 
SSC $\nu F_\nu$ flux, 
\begin{equation}
 f_\e^{SSC} = {3\over 4} c \sigma_{\rm T} \e^{\prime 2}_s\;{\dD^4\over 4\pi d_L^2}\;
\int_0^\infty d\ep\;{u^\prime (\ep )\over \e^{\prime 2}}
\int_{\gp_{min}}^{\gp_{max}}d\gp\;{N_e^\prime(\gp )\over \g^{\prime 2}}\;F_{\rm C} (q^\prime,\Gamma^\prime_e)\;,
\label{epsjssc}
\end{equation}
where
\begin{equation}
F_{\rm C}(q^\prime,\Gamma^\prime_e
)= \left[ 2q^\prime \ln q^\prime +(1+2q^\prime)(1-q^\prime) +{1\over 2}
{(\Gamma^\prime_e q^\prime)^2\over (1+\Gamma^\prime_e
 q^\prime)}(1-q^\prime) \right]
\;H\;\left( q^\prime; {1\over 4\gamma^{\prime 2}},1\right),
\label{fcq}
\end{equation}
\begin{equation}
q^\prime \equiv {\ep_s/\gp \over \Gamma^\prime_e
(1-\ep_s/\gp )}\;,\;{\rm and}\; \; \Gamma^\prime_e = 4\ep\gp\;.
\label{jesCq}
\end{equation}
The synchrotron photons provide a target radiation field with
spectral energy density
\begin{equation}
u^\prime (\ep ) = \epsilon^{\prime}m_ec^2 n^\prime_{syn}(\ep ) =  
\frac{ 3 d_L^2 f_\e^{syn} }{ cR_B^{\prime 2}\dD^4\ep }\;,
\label{uprime}
\end{equation}
using eq.\ (\ref{gps}).  
The scattered photon energy in the comoving frame is related to the observed photon
energy $h\nu = m_ec^2 \e$ by the relation 
\begin{equation}
\ep_s \;=\; {(1+z)\e\over \dD }\; \equiv\; {\e_s\over \dD}\;.
\label{eps}
\end{equation}
From the limits on the integration over $\gp $ implied by the limits on $q^\prime$ we find
\begin{equation}
\gp_{min} = {1\over 2} \ep_s\;\left( 1+\sqrt{1+{1\over \ep\ep_s}} \;\right)
\label{gpmin}
\end{equation}
and
\begin{equation}
\gp_{max} = {\ep\ep_s\over \ep - \ep_s}H(\ep -  \ep_s) \;+\; \gp_2H(\ep_s - \ep )\;
\label{gpmax}
\end{equation}
\citep[see][for a detailed derivation of synchrotron/SSC models 
and application to  blazars]{fdb08}.
Here the maximum lepton Lorentz factor injected 
into the radiating fluid is $\gp_2$, and the Heaviside
function $H(x;a,b)$ is defined such that $H(x;a,b) = 1$ when $a\leq x \leq b$, and 
$H(x;a,b) = 0$ otherwise; the Heaviside function
 with one entry is defined such that $H(x) = 1$ when $x\geq 0$, and $H(x) = 0$
otherwise.
The $\nu F_\nu$ SSC spectrum is therefore given by
\begin{equation}
f_\e^{SSC} = \left( {3\over 2}\right)^3\; {d_L^2 \e_s^{\prime 2}
\over 
R_b^{\prime 2} c \dD^4 U_B}\;\int_0^\infty d\ep\;{f_{\tilde \e}^{syn}\over \e^{\prime 3}}\;
\int_{\gp_{min}}^{\gp_{max}}d\gp\;{F_{\rm C}(q^\prime,\Gamma^\prime_e)f_{\hat \e}^{syn}\over \g^{\prime 5} }\;,
\label{feSSC1}
\end{equation}
where $\tilde \e \equiv {\dD \ep/ (1+z)}$ and 
$\hat \e \equiv {\dD \e_B \g^{\prime 2}/ (1+z)}$.
The maximum $\nu F_\nu$ SSC flux at photon energy
$\e^{SSC}_{pk}$ can be approximated in the Thomson limit by the expression
\begin{equation}
\label{fssc}
f_{\e_{pk}^{SSC}}^{SSC} \simeq \frac{24\pi d_L^2 (1+z)^2}{(\dD^3Bt_v)^2 c^3}\ 
	\left( f_{\e_{pk}^{syn}}^{syn} \right)^2\;,
\end{equation}
where
the peak frequencies are related by
\begin{equation}
\e_{pk}^{syn} \;\cong \; \sqrt{ {\e_{pk}^{SSC}\e_B\dD\over 1+z}}\;
\label{essc}
\end{equation}
\citep{tav98,fdb08}. 
Here $f_{\e_{pk}^{syn}}^{syn}$ is the $\nu F_\nu$ peak of the synchrotron 
component, which reaches its maximum at $\e = \e_{pk}^{syn}$. 

Second-order SSC takes place when the SSC photons are again Compton
scattered by electrons in the same blob, and may account for
superquadratic variability of the $\gamma$-ray flux with respect to
the synchrotron flux \citep{per08}.  In principle, these photons can
again be Compton scattered to arbitrarily higher orders, though
higher-order scatterings are negligible due to Klein-Nishina effects.
Calculating second-order SSC can be done accurately by replacing

$f_{\tilde \e}^{syn}$ with $f_{\tilde \e}^{SSC}$ in eq. (\ref{feSSC1}), so 
that
\begin{equation}
f_\e^{SSC,2} = \left( {3\over 2}\right)^3\; {d_L^2 \e_s^{\prime 2}
\over 
R_b^{\prime 2} c \dD^4 U_B}\;\int_0^\infty d\ep\;{f_{\tilde \e}^{SSC}\over \e^{\prime 3}}\;
\int_{\gp_{min}}^{\gp_{max}}d\gp\;{F_{\rm C}(q^\prime,\Gamma^\prime_e)f_{\hat \e}^{syn}\over \g^{\prime 5} }\;.
\label{feSSC2ndorder}
\end{equation}

\subsection{ $\gamma\gamma$ Opacity}

Gamma-ray photons  are subject
to $\g\g$ attenuation by synchrotron photons produced in the 
radiating plasma, by ambient photons in the environment of 
the black hole (starred frame), and  by 
photons of the intergalactic radiation field.
The $\g\g$ pair-production cross section 
\begin{equation}
 \sigma_{\gamma\gamma} (s)={1\over 2}\pi r_e^2  (1-\bcm^2)
\left[(3-\bcm^4)\ln\left({1+\bcm\over 
1-\bcm}\right) - 2\bcm (2-\bcm^2)\right]\;
\label{sigmagg}
\end{equation}
\citep{jr76,nik61,gs67,bmg73}, 
where $\gcm$ is the center-of-momentum
frame Lorentz factor of the produced electron and positron,
$\bcm = (1-\gcm^{-2})^{1/2} = \sqrt{1-s^{-1}}$,
\begin{equation}
s = \gcm^2 = {\e_*\e_1(1+z)\over 2} (1-\cos\psi) \;,
\label{scmenergy}
\end{equation}
and $r_e =e^2/m_ec^2 
\cong 2.8179\times 10^{-13}$ cm
is the classical electron radius. The interaction
angle $\psi$, given by the relation 
\begin{equation}
\cos\psi = \mu_* \mu_s + \sqrt{1-\mu_*^2}\sqrt{1-\mu_s^2}\cos(\phi_*-\phi_s)\;,
\label{cospsi}
\end{equation}
is the angle between the directions of the photon detected with energy $\e_1$
 and the target photon with
energy $\e_*$. 

The absorption probability per unit pathlength is
\begin{equation}
{d\tau_{\g\g}(\e_1)\over dx} = 
\oint d\Omega_* \;(1-\cos\psi) \int_0^\infty d\e_*
\;n_{ph}(\e_*,\Omega_*)\;\sigma_{\g\g}(s)\;.
\label{dtauggdx}
\end{equation}
For absorption by synchrotron photons
within the radiating volume, $\e_* \rightarrow \ep$ and $\e_1\rightarrow \ep_1 =
(1+z)\e_1/\dD$, and the target synchrotron 
radiation field is given by eq.\ (\ref{uprime}). In this case, 
the optically-thin $\gamma$-ray emission spectrum is modified by
the factor  $3u(\tgg) / \tgg$ for a spherical geometry,
where
\begin{equation}
u(\tgg) = {1\over 2} + {\exp(-\tgg )\over \tgg} - {1-\exp(-\tgg )\over \tgg^2} \;.
\label{utgg}
\end{equation}
Here $\tau_{\g\g}$ is
the total $\g\g$ optical depth integrated over pathlength.
For absorption by ambient photons in the vicinity of the AGN, $\e_*$ is 
the photon energy in the AGN rest frame. For cosmic $\g\g$ absorption, the target photons are given by the 
spectrum of the intergalactic background light, 
which evolves with redshift. In the latter two cases, the intrinsic spectrum
is modified by the factor $\exp(-\tau_{\g\g})$.

\section{Compton-Scattered External Radiation Fields}

In the one-zone model, the $\nu F_\nu$ spectrum of Compton-scattered
external radiation fields 
is given by the Compton spectral 
luminosity $\e_s L_{\rm C}(\e_s,\Omega_s)$  according to the relation
\begin{equation}
f^{\rm C}_{\epsilon} = {\e_s L_{\rm C}(\e_s,\Omega_s)\over d_L^2}\;,
\label{fecompton}
\end{equation}
where $\e_s \equiv (1+z)\e$, from eq.\ (\ref{eps}), and 
$\Omega_s = \Omega$. The latter equality 
means that the photon direction is not deflected
in transit to the observer.
The Compton spectral luminosity is given by
$$\e_s L_{\rm C}(\e_s,\Omega_s )= m_ec^3\e_s^2\oint d\Omega_* \int_0^\infty d\e_* \;n_{ph}(\e_*,\Omega_* )
\oint d\Omega_e \int_1^\infty d\g\; N_e(\g ,\Omega_e)\times $$
\begin{equation}
\;(1-\cos\psi )\;{d\sigma_{\rm C}(\bar\e)\over d\e_s}\;
\delta (\Omega_s - \Omega_e )\;,
\label{esJesOs}
\end{equation}
having already introduced the approximation that the scattered photon travels
in the same direction as the relativistic scattering electron, i.e., 
$\Omega_s = \Omega_e$. Because of this approximation,
the cosine of the angle $\psi$ is given by eq.\ (\ref{cospsi}). The invariant collision energy 
\begin{equation}
\bar\e \equiv \gamma \e_* (1-\sqrt{1-1/\g^2}\cos\psi)\cong \gamma \e_* (1-\cos\psi)\;
\label{bare}
\end{equation}
because $\gamma \gg 1$. The relation $n_{ph}(\e_*,\Omega_*) =
u(\e_*,\Omega_*)/(m_ec^2\e_*)$ gives the specific spectral number
density of target photons with energy $\e_*$, the starred quantities
referring to the frame stationary with respect to the black hole.

The Compton cross section in the head-on approximation is given by
\begin{equation}
{d\sigma_{\rm C}\over d\e_s} \;\cong\;
{\pi r_e^2 \over \gamma\bar\e}\;\Xi\;H\left(\e_s ;
{\bar\e\over 2\gamma}, {2\gamma\bar\e\over 1+2\bar\e}\right)\,
\label{dsigKNdedO_1}
\end{equation}
\citep{ds93,db06},
where 
\begin{equation}
\Xi \;\equiv \;
y+y^{-1}  - {2\e_s\over \gamma \bar\e y} + 
({\e_s\over \gamma \bar\e y})^2 \;,
\label{Xi}
\end{equation}
\begin{equation}
y \;\equiv\; 1 - {\e_s\over\g} \;,
\label{y}
\end{equation}
$\bar \e$ is given by eq.\ (\ref{bare}).
The Compton spectral luminosity in the head-on approximation becomes
\begin{equation}
\e_s L_{\rm C}(\e_s,\Omega_s) = {c\pi r_e^2 \e_s^2 }\oint d\Omega_*\; \int_0^{\e_{*,hi}} 
d\e_*\; {u_*(\e_*,\Omega_* )\over \e_*^2}
\int_{\g_{low}}^\infty
d\g\;\g^{-2} N_e(\g,\Omega_s )\; 
\Xi\;.
\label{esLCesomegas}
\end{equation}
The lower limit on the electron Lorentz factor $\g_{low}$ and the upper limit $\e_{*,hi}$ 
implied by the kinematic limits on $y$ are 
\begin{equation}
\g_{low} = {\e_s\over 2}\;\left[ 1 + \sqrt{1 + {2\over \e_*\e_s (1-\cos\psi)}}\;\right]\;,
\label{glow}
\end{equation}
and
\begin{equation}
\e_{*,hi} = {2\e_s\over 1-\cos\psi}\;.
\label{ehi}
\end{equation}

Eq.\ (\ref{esLCesomegas}) is the starting point to calculate accurate 
Compton-scattered spectra involving relativistic electrons 
and external photon fields with arbitrary
anisotropies. In contrast to the comoving electron spectrum used in the SSC calculation, 
the calculation of Compton-scattered radiation uses 
the electron spectrum $N_e(\g,\Omega_e)$ and the target photon spectrum 
defined in the stationary frame \citep{gkm01}.
The invariant phase volume 
$d{\cal V} =  dV d^3\vec p$ for
relativistic particles is given by 
\begin{equation}
{dN\over d{\cal V}} = {dN\over dV d^3\vec p} \;=\;{1\over (m_ec)^3}\;{1\over
\gamma^2}\; {dN\over d\gamma d\Omega dV}\;=\;inv\;,
\label{dNdV}
\end{equation}
implying that
\begin{equation}
 N_e(\gamma, \Omega) =
{\gamma^2\over \gamma^{\prime 2}}\;{dV\over dV^\prime}\;
N_e^\prime(\gamma^\prime,\Omega^\prime ) = \dD^3
N_e^\prime(\gp,\Omega^\prime )\;,
\label{PdN/dV_1}
\end{equation}
noting that $dV/dV^\prime = dt^\prime/dt = \dD$, and $\gamma = 
\dD \gp$ when $\gp, \g \gg 1$, required for the
head-on approximation. 
For an isotropic comoving distribution of electrons, 
$N_e(\g,\Omega_s) = \dD^3 N^\prime_e(\gp )/4\pi$. Hence
\begin{equation}
\e_s L_{\rm C}(\e_s,\Omega_s) = {cr_e^2  \over 4}\;\e_s^2\dD^3\;\int_0^{2\pi}d\phi_*
\int_{-1}^1 d\mu_*\; \int_0^{\e_{*,hi}} d\e_*\; {u_*(\e_*,\Omega_* )\over \e_*^2}\int_{\g_{low}}^\infty
d\g\;\g^{-2} {N^\prime_e(\g/\dD )}\; 
\Xi\;, 
\label{esLCesomegas_1}
\end{equation}
%and $\e_{hi} = 2\e_s/(1-\cos\psi )$.
or
\begin{equation}
f_\e^{\rm EC} = {\e_s L_{\rm C}(\e_s,\Omega_s)\over d_L^2} 
= {c\pi r_e^2  \over 4\pi d_L^2}\;\e_s^2\dD^3\;\int_0^{2\pi}d\phi_*
\int_{-1}^1 d\mu_*\; \int_0^{\e_{*,hi}} d\e_*\; {u_*(\e_*,\Omega_* )\over \e_*^2}\int_{\g_{low}}^\infty
d\g\; {N^\prime_e(\g/\dD )\over \g^2}\; 
\Xi\;.
\label{esLCesomegas_26}
\end{equation}

In terms of the measured synchrotron $\nu F_\nu$ spectrum, eq.\ (\ref{Neprimegps}),
the source Compton spectrum for external Compton (EC) scattering in a standard
one-zone model for blazars is, in general,
given by the four-fold integral
\begin{equation}
f_\e^{\rm EC}  = \left( {3\over 4}\right)^2\;{\e_s^2 \dD^2\over U_B}\;\int_0^{2\pi}d\phi_*
\int_{-1}^1 d\mu_* \int_0^{\e_{hi}} d\e_*\; {u_*(\e_*,\Omega_* )\over \e_*^2}\int_{\g_{low}}^\infty
d\g\; {f^{syn}_{\breve\e}\over \g^5}\; 
\Xi\;\;,
\label{esLCesomegas_2}
\end{equation}
with
\begin{equation}
\breve\e \equiv {\e_B\g^2\over (1+z)\dD}\;,
\label{brevee}
\end{equation}
using eq.\ (\ref{Neprimegps}). The number of integrations can 
obviously be reduced by choosing symmetrical target photon geometries.

\subsection{Point Source Radially Behind Jet}

First we consider the flux when nonthermal electrons
in a relativistic jet Compton-scatter photons from a
point source of radiation, isotropically emitting and 
located radially behind the outflowing 
plasma jet. For a monochromatic point source with luminosity $L_0$
and energy $\e_0$, 
the spectral luminosity can be expressed as
\begin{equation}
L_*(\e_*)  \; =\;  L_0 \delta(\e_* - \e_0 )\;.
\label{Lstarestar}
\end{equation}
The spectral energy distribution of the target photon 
source at distance $r$ from the point source is therefore given by
\begin{equation}
u(\e_*,\Omega_* ) = {L_0\over 4\pi r^2 c}\;{\delta(\mu_* - 1)\over 2\pi }\;\delta(\e_* - \e_0 )\;.
\label{ueomega}
\end{equation} 
Substituting eq.\ (\ref{ueomega}) into eq.\ (\ref{esLCesomegas_1}) 
and solving gives
\begin{equation}
\e_s L^{pt}_{\rm C}(\e_s,\Omega_s) = {r_e^2 \e_s^2 L_0 \dD^3\over 16\pi r^2 \e_0^2}
\int_{\bar\g_{low}}^\infty
d\g\;{N^\prime_e(\g/\dD )\over \g^2}\; 
\bar\Xi\;.
\label{esLCesomegas_3}
\end{equation}
Using eq.\ (\ref{fecompton}), eq.\ (\ref{esLCesomegas_3}) becomes
\begin{equation}
f^{{\rm C},pt}_\e = 
{r_e^2 \e_s^2 L_0 \dD^3\over 16\pi r^2d_L^2 \e_0^2}
\int_{\bar\g_{low}}^\infty
d\g\;{N^\prime_e(\g/\dD )\over \g^2}\; 
\bar\Xi\;,
\label{feptNegamma17}
\end{equation}
or with eq.\ (\ref{Neprimegps}),
\begin{equation}
f^{{\rm C},pt}_\e = {3^2\over 8^2\pi}\; 
{L_0 \e_s^2 \dD^2\over c r^2  U_B \e_0^2}\;
\int_{\bar\g_{low}}^\infty d\g\;{f_{\breve\e}^{syn}\over \g^5}\;
\bar\Xi\;,
\label{fept}
\end{equation}
where $\breve \e$ is defined in eq.\ (\ref{brevee}), $\bar\Xi$ is defined 
by eq.\ (\ref{Xi}) with $\bar\e$ replaced by $\bar{\bar\e} = \g \e_0(1-\mu_s)$,
and
\begin{equation}
\bar\g_{low} = {\e_s\over 2}\;\left[ 1 + \sqrt{1 + {2\over \e_0\e_s (1-\mu_s)}}\;
\right]\;.
\label{barglow}
\end{equation}

Eqs.\ (\ref{feptNegamma17}) and (\ref{fept}) give the Compton-scattered
spectrum from a point source of radiation located radially behind the jet, 
generalizing the Thomson-regime result \citep{dsm92} to include scattering
in the Klein-Nishina regime.
A scattered disk component should be found in all blazar models, with 
its importance strongly dependent on distance $r$ of  the jet from the accretion 
disk. The point-source approximation gives the least upscattered
flux in the Thomson limit, and an extended disk having the same power as a point source
will give a more intense flux. At sufficiently large jet heights $r \gg \Gamma^4 R_g$,
defining the far field, 
where $R_g = GM/c^2
\cong 1.5\times 10^{13} M_8$ cm is the gravitational radius, 
the Shakura-Sunyaev disk can be described as a point source 
radially behind the the jet. Photons from large disk radii are important
in the near field $r \ll \Gamma^4 R_g$ \citep{ds02}.

\subsubsection{Reduction to the Thomson Regime}

We now derive the Thomson limit for the 
$\nu F_\nu$ spectrum, eq.\ (\ref{feptNegamma17}). Because we consider 
relativistic electrons $\g \gg 1$, we are restricted to the condition
$\bar\gamma_{low} \gg 1$, which occurs according to eq.\ (\ref{barglow}) 
 when 
either $\e_s \gg 1$ or
$\e_s/\e_0(1-\mu_s) \gg 1$. The Thomson condition can be expressed as
$\e_s \ll \gamma$, which is guaranteed when $\e_s \ll \bar\g_{low}$, in which 
case $2\e_0\e_s (1-\mu_s) \ll 1$. Another statement of the Thomson condition
is that $\gamma\e_0 (1-\mu_s)\ll 1$ which, with $\e_s \ll \g$, again implies that
$\e_0\e_s(1-\mu_s) \ll 1$. Thus
\begin{equation}
\bar\g_{low} \rightarrow \sqrt{\e_s\over 2\e_0(1-\mu_s)}\;.
\label{barglow1}
\end{equation}

For the scattering kernel, eq.\ (\ref{Xi}), $\e_s\ll \gamma$ and $y \rightarrow 1$
in the Thomson regime, so 
\begin{equation}
\bar \Xi \rightarrow  \bar \Xi_{\rm T} \equiv 
2 - 2\left({\e_s\over \g\bar{\bar \e}}\right) + \left({\e_s\over \g \bar{\bar \e}}
\right)^{2}\;.
\label{XiT}
\end{equation}
Away from the endpoints of the spectrum, $\e_s\ll \g \bar{\bar \e}$ and 
$\bar \Xi_{\rm T} \rightarrow  2$. Hence
\begin{equation}
f_\e^{pt,{\rm T}} \; = \; {3\over 4} \;{\sT L_0\over (4\pi r d_L)^2}\;
\left({\e_s\over \e_0}\right)^2\dD^3\;\int_{\dD\bar{\bar\g }}^\infty d\g 
\; {N^\prime_e(\gp )\over \g^2}\;,
\label{feptT}
\end{equation}
defining 
\begin{equation}
\dD\bar{\bar\g } \equiv \sqrt{ {\e_s \over 2\e_0(1-\mu_s)} }\;.
\label{dDbarbarg}
\end{equation}
For the comoving electron distribution, eq.\ (\ref{Neprimegps}),  in the 
power-law form
\begin{equation}
N_e^\prime(\gp ) = K^\prime \g^{\prime -p}H(\gp;\gp_1,\gp_2)\;,
\label{Neprime}
\end{equation}
eq.\ (\ref{feptT}) becomes 
$$f_\e^{pt,{\rm T}} \; = \; {3\over 4(p+1)} \;{\sT L_0 K^\prime\over (4\pi r d_L)^2}\;
\left({\e_s\over \e_0}\right)^2\dD^{3+p}\;\left[\max \left( \bar{\bar\g },
\gp_1\right)^{-(p+1)} - \g_2^{\prime -(p+1)} \right]
\;$$
\begin{equation}
\rightarrow  {3\over 4(p+1)} \;{\sT L_0 K^\prime\over (4\pi r d_L)^2}\;
\left({\e_s\over \e_0}\right)^{(3-p)/2}\dD^{3+p}\; [2(1-\mu)]^{(p+1)/2}
\;,
\label{feptT_1}
\end{equation}
where the final expression applies in the regime $\gp_1 \ll \bar{\bar\g }\ll \gp_2$.
This can 
be written as 
\begin{equation}
f_\e^{pt,{\rm T}}\cong  \left({3\over p+1}\right)\; \dD^6 (1-\mu_s)^2\;\left({\sigma_{\rm T} \over 4\pi d_L^2}\right)
\;\left( {L_0\over 4\pi r^2}\right) \;\bar {\bar\g}^{3}N^{\prime}_e(\bar{\bar \g})\;,
\label{feptT_2}
\end{equation}
which can be compared with the Thomson-regime expression 
\begin{equation}
f_\e^{pt,{\rm T}} \cong \left({1\over 2}\right)\;\dD^6 (1-\mu_s)^2\;\left({\sigma_{\rm T} \over 4\pi d_L^2}\right)
\;\left( {L_0\over 4\pi r^2}\right) \;\tilde \g^{\,\prime 3}N^{\prime}_e(\tilde \gp)\;
\label{feptT_dsm}
\end{equation}
\citep{dsm92,ds93,ds02}, where
\begin{equation}
\tilde \gp = {1\over \dD}\;\sqrt{\e(1+z)\over \e_0(1-\mu_s)} = \sqrt{2} \bar{\bar \g}\;.
\end{equation}

\subsubsection{Accurate Thomson Regime Spectrum}

Eq.\ (\ref{feptT_2}) for the Thomson-scattered
spectrum was derived assuming  $\Xi_{\rm T}=2$, 
away from the endpoints of the spectrum. Using the full expression for $\Xi_{\rm T}$
and a power-law electron distribution gives an accurate expression for the 
spectrum  of a localized jet of isotropically entrained 
electrons Thomson scattering a point source, monochromatic
radiation field that enters the jet from behind. The result is
\begin{equation}
f_\e^{pt,{\rm T}}=  \;
{\pi r_e^2 L_0 \over (4\pi rd_L)^2}\;
 \big({\e_s\over\e_0}\big)^2 \dD^3 \;\int_{\dD\bar{\bar \g}}^\infty
d\g\; \big( {2\over \g^2} - {2s\over \g^4} +{s^2\over \g^6}\big) 
 N^{~\prime}_e(\gp)\;,
\label{feptT_3}
\end{equation}
where
\begin{equation}
s \;\equiv\; {\e_s\over \e_0(1-\mu_s)} \; = \; 2 \dD^2 {\bar{\bar \g}}^2\;.
\end{equation}

For the power-law electron distribution $N_e^\prime(\gp )$, eq.\ (\ref{Neprime}), the accurate 
analytic Thomson-regime $\nu F_\nu$ flux from an isotropic monochromatic point source 
radiation field located behind the jet is 
$$f_\e^{pt,{\rm T}}=  {3\over 4}\; {\sigma_{\rm T} L_0 K^\prime\over (4\pi r d_L)^2}\;
\big({\e_s\over \e_0}\big)^2\;\dD^{3+p}\;\big\{ {1\over 1+p}\big[\g_1^{-(1+p)}-\g_2^{-(1+p)}\big]
$$
\begin{equation}
-\;{s\over (3+p)}\big[\g_1^{-(3+p)}-\g_2^{-(3+p)}\big]\;
+
{s^2\over 2(5+p)}\big[\g_1^{-(5+p)}-\g_2^{-(5+p)}\big]\big\},
\label{feptT_4}
\end{equation}
where $\g_1 = \max(\dD\gp_1,\dD{\bar{\bar \g}})$ and $\g_2 = \dD\gp_2$.

In the asymptotic limit $\gp_1 \ll \bar{\bar \g }\ll \gp_2$, 
eq.\ (\ref{feptT_4}) approaches
\begin{equation}
f_\e^{pt,{\rm T}}\rightarrow {\cal A}(p) \;\big({\sigma_{\rm T} \over 4\pi d_L^2}\big)
\;\big( {L_0\over 4\pi r^2}\big)
\;\dD^6 (1-\mu_s)^2\;
\bar {\bar\g}^{3}N^{~\prime}_e(\bar{\bar \g})\;,
\label{feptT_5}
\end{equation}
where
\begin{equation}
{\cal A}(p) \equiv 3\;\left({1\over 1+p} - {2\over 3+p} + {2\over 5+p}\right)\;.
\label{cala}
\end{equation}
The values of ${\cal A}(p) = 1.0, 0.657$, and $ 0.5$ for $p = 1, 2$, and $3$, 
respectively. 
For the Thomson approximation away from the 
endpoints of the spectrum, given by eq.\ (\ref{feptT_2}), the corresponding coefficient 
is $3/(p+1)$.

\subsubsection{Solution with Compton Cross Section}

From eqs.\ (\ref{esLCesomegas_3}) and (\ref{fecompton}), 
\begin{equation}
f^{pt,{\rm C}}_\e = {\pi r_e^2 L_0 \over (4\pi r d_L)^2}\;\left({\e_s\over \e_0^2}\right)
\;\dD^3
\int_{\bar\g_{low}}^\infty
d\g\;{N^\prime_e(\g/\dD )\over \g^2}\; \left[y+y^{-1}  - {2\e_s\over \gamma \bar\e y} + 
({\e_s\over \gamma \bar\e y})^2\right]\;\;.
\label{esLCesomegas_4}
\end{equation}
Introducing
$u = {\e_s/ \g }$ and $v = \e_s\e_0(1-\mu_s)\;$
and  changing variables to $u$ gives, for the power-law electron distribution,
eq.\ (\ref{Neprime}), 
\begin{equation}
f^{pt,{\rm C}}_\e = {\pi r_e^2 L_0 K^\prime\over (4\pi r d_L)^2}\;
\left({\e_s\over \e_0^2}\right)
\;\dD^{3+p}\e_s^{-(1+p)}I_{\rm C}\;,
\;
\label{esLCesomegas_5}
\end{equation}
where
\begin{equation}
I_{\rm C}  = \int_{u_1}^{u_2}
du\;\left[ \;
u^p - u^{p+1} +{u^p\over 1-u} - {2u^{p+2}\over v(1-u)}
+{u^{p+4}\over v^2(1-u)^2}\;
\right]\;,
\label{ic1}
\end{equation}
and $u_1 = \e_s\dD/\gp_2$, $u_2 = \e_s\min\left({1\over \bar\g_{low}},{\dD\over \gp_1}\right)$.
The series solution of eq.\  (\ref{ic1}) is given by
\begin{equation}
I_{\rm C}  = \left\{ {u^{p+1}\over p+1} - {u^{p+2}\over p+2} +
\sum_{i=0}^\infty \left[ {u^{i+p+1}\over i+p+1}-{2u^{i+p+3}\over v(i+p+3)}
+{1\over v^2}{(i+1) u^{i+p+5}\over i+p+5}
\right]\right\} \Biggr|_{u_1}^{u_2}\;.
\label{ic2}
\end{equation}
Eq.\ (\ref{ic1}) can be solved analytically for integral $p$.
%The analytic expressions for $I_{\rm C}(p)$ for $p = 1,$ 2, and 3 are given in the Appendix.

%%%%%%%%%%%%%%%%%%%%%%%%%%%%%%%%%%%%%%%%%%%%%%%%%%%%%%%%%%%%%%%%%%%%%
%%%%%%%%%%%%%%%%%%%%%%%%%%%%%%%%%%%%%%%%%%%%%%%%%%%%%%%%%%%%%%%%%%%%%

\subsection{Shakura-Sunyaev Accretion Disk Field}

For the emission spectrum of an accretion disk surrounding the 
supermassive black hole, we consider the cool, optically-thick 
blackbody solution of \citet{ss73}. The
disk emission is approximated by a surface radiating at the blackbody
temperature associated with the local energy dissipation rate per unit
surface area, which is derived from considerations of viscous
dissipation of the gravitational potential energy of the accreting
material. The accretion luminosity is defined 
in terms of the Eddington ratio
\begin{equation}
\ell_{\rm Edd} = {\eta\dot m c^2\over L_{\rm Edd}}\;,
\label{ellEdd}
\end{equation}
where $\eta\sim 0.1$ is the efficiency to transform accreted matter to
escaping radiant energy.  The Eddington luminosity $L_{\rm Edd}=
1.26\times 10^{46} M_8$ ergs s$^{-1}$, where the mass of the central
supermassive black hole is $M =10^8M_8 M_\odot$ and the black hole is
accreting mass at the rate $\dot m$ (gm s$^{-1}$).
 
For steady flows where the energy is derived from the viscous dissipation
of the gravitational potential energy of the accreting matter, the
radiant surface-energy flux
\begin{equation}
{d{\cal E}\over dAdt}  = 
{3GM\dot m\over 8\pi R^3}\; \varphi(R)
\label{PhiE}
\end{equation}
\citep{ss73}, where 
\begin{equation}
\varphi(R) =[1-\beta_i(R_i/R)^{1/2}]\;,
\label{varphiSchwarzschild}
\end{equation}
 $\beta_i \cong 1$, and $R_i = 6GM/c^2$ for the Schwarzschild
metric. Integrating equation (\ref{PhiE}) over a two-sided disk gives
$\eta = 1/12$.  Assuming that the disk is an optically-thick
blackbody, the effective temperature of the disk can be
determined by equating equation (\ref{PhiE}) with the surface energy
flux $\sigma_{\rm SB}T^4(R)$.
A monochromatic approximation for the mean
photon energy $m_ec^2\bar \e (R) = k_{\rm B} T(R)$ at radius
$R$ of the accretion disk with mean temperature 
$T(R)$ is given by
\begin{equation}
m_ec^2\langle \e (R)\rangle \;\cong \;2.70{\kB T(R)} \; \cong \;2.70 {\kB }
\big[ {3GM\dot m\varphi(R)\over 8\pi
R^3\sigma_{\rm SB}}\big ]^{1/4}\;\cong 
137\;\big({\ell_{\rm Edd}\over M_8\eta}\big)^{1/4}\tilde R^{-3/4}\;{\rm eV}
\;, 
\label{<e>}
\end{equation}
so
\begin{equation}
\langle \e (\tilde R)\rangle \;\cong 
\;2.7\times 10^{-4}\;\xi \tilde R^{-3/4}
\;,
\label{<e>_1}
\end{equation}
where 
\begin{equation}
\xi  \equiv \left({\ell_{\rm Edd}\over 
M_8 \eta}\right)^{1/4}\;.
\label{xidef1}
\end{equation}
Here $\theta = \arccos \mu_*$ is the angle between the directions of the jet and the photon 
that intercepts the jet, 
\begin{equation}
\tilde R = {R\over R_g} = \tilde r\sqrt{\mu_*^{-2}-1}\;,
\end{equation} 
where $\tilde r = {r/ R_g}$. 
Lengths marked with a tilde are normalized to $R_g$. The final expression in eq.\ (\ref{<e>}) is 
valid when $\tilde R \gg \tilde R_{min}$, though we assume it is reasonably accurate to $\tilde R\cong
\tilde R_{min}$.

The intensity of the Shakura-Sunyaev accretion disk model along 
the jet axis is
given by
\begin{equation}
I_\e^{\rm SS}(\Omega;R) \cong \;{3GM\dot m\over 16\pi^2
R^3}\varphi(R)\; \delta[\e - {2.7 k_{\rm B}\over m_ec^2}\;T(R)]\;
\label{uoverc}
\end{equation}
\citep{st83}. Thus 
\begin{equation}
I_\e^{\rm SS}(\Omega;\tilde R) = 
{3\over 16\pi^2}\;{\ell_{\rm Edd}L_{\rm Edd}\over \eta \tilde R R^2}\;
\varphi(\tilde R)\delta [\e -\langle\e(\tilde R)\rangle ]\;
\label{IeSS}
\end{equation}
\citep{ds02}.
Substituting eq.\ (\ref{IeSS}) into eq.\ (\ref{esLCesomegas_1}), using the 
relation $I_\e (\Omega)  = c u(\e ,\Omega)$, gives
$$f_\e^{\rm SS} = {3^2\over 2^9\pi^3}\;{\sT\e_s^2\over d_L^2 R_g^2}\;
{ \ell_{\rm Edd} L_{\rm Edd}\over\eta \tilde r^3 }\; \dD^3\;\int_0^{2\pi}d\phi_*
\int_{\mu_{*,min}}^{\mu_{*,max}} d\mu_*\;  \;{\varphi(\tilde R )\over (\mu_*^{-2}-1)^{3/2}
\langle\e(\tilde R)\rangle^2}$$
\begin{equation}
\times\int_{\bar\g_{low}}^\infty
d\g\;\g^{-2} {N^\prime_e(\g/\dD )}\; 
\big[y+y^{-1}  - {2\e_s\over \gamma \tilde{\bar\e} y} + 
({\e_s\over \gamma \tilde{\bar\e} y})^2 \big]\;.
\label{eslcssexyz}
\end{equation}
The integration over angle in eq.\ (\ref{eslcssexyz}) 
is limited by the inner radius of the accretion 
disk, so that 
$\mu_{*,max} = \big[1 + (6/\tilde r)^2\big]^{-1/2}$ for a Schwarzschild black hole,
and
\begin{equation}
\tilde{\bar\e} = \tilde{\bar\e}(\g,\e_s,\psi) =\g \langle\e(\tilde R)\rangle(1-\cos\psi )\;,
\label{tildebare}
\end{equation}
and 
\begin{equation}
\bar\g_{low} = {\e_s\over 2}\;\left( 1 + \sqrt{1+{2\over \langle\e(\tilde R)\rangle\e_s(1-\cos\psi)}}\right)\;. 
\end{equation}
The other limit on the angular integration arises because of the restriction given 
by eq.\ (\ref{ehi}), so that
\begin{equation}
\langle \e (\tilde R )\rangle \; < \; {2\e_s\over 1-\cos\psi }\;,
\label{limit}
\end{equation}
which restricts the integral to a maximum value of $\tilde R$ and therefore 
$\mu_* \gtrsim \mu_{*,min}$.
In the calculation of 
$\cos\psi$, eq.\ (\ref{cospsi}), we take $\phi_s = 0$ without loss of generality because of the 
assumed azimuthal symmetry of the accretion-disk emission.

The result for the accretion-disk radiation field scattered
by isotropic, relativistic jet electrons is a 3-fold integral---reduced
from a 4-fold integral by approximating the disk blackbody spectrum by its mean
thermal energy at different radii. 
When expressed in terms of the measured synchrotron $\nu F_\nu$
spectrum using eq.\ (\ref{Neprimegps}), the result for the 
accretion-disk radiation field is
$$f_\e^{\rm SS} = {3^3\over 2^8\pi^2}\;{\e_s^2\over c R_g^2 U_B}\;
{ \ell_{\rm Edd} L_{\rm Edd}\over\eta \tilde r^3 }\;\dD^2
\;\int_0^{2\pi}d\phi_*
\int_{\mu_{*,min}}^{\mu_{*,max}} d\mu\;  \;{\varphi(\tilde R )\over (\mu_*^{-2}-1)^{3/2}
\langle\e(\tilde R)\rangle^2}$$
\begin{equation}
\times\int_{\bar\g_{low}}^\infty
d\g\;\g^{-5} {f^{syn}_{\breve\e}}\; 
\left[y+y^{-1}  - {2\e_s\over \gamma \tilde{\bar\e} y} + 
({\e_s\over \gamma \tilde{\bar\e} y})^2 \right]\;,
\label{eslcssexyza}
\end{equation}
recalling the definitions of $\gp_s$ and $\breve \e$ from 
eqs.\ (\ref{gps}) and (\ref{brevee}), respectively.
This can be reduced to a 2-fold
integral by approximating a typical scattering as occurring at $\phi_* = \pi/2$, 
so that $\tilde{\bar\e} = \gamma\e (1-\mu_*\mu_s)$. 
Because it is  feasible to perform the 3-fold integral numerically, we show
results of the more accurate calculations.

%%%%%%%%%%%%%%%%%%%%%%%%%%%%%%%%%%%%

\subsubsection{Regimes in Compton-Scattered Accretion-Disk Spectra}

\label{trends}

The limiting behaviors of the Compton scattered spectra can be understood 
based on simple $\delta$-approximations in the Thomson regime.  
 External Compton scattering of the Shakura-Sunyaev
disk with radiant luminosity $L_d = \ell_{\rm Edd} L_{\rm Edd}$
 in the 
near field, i.e., $\tilde r\ll \Gamma^4$, can be approximated as 
\begin{equation}
\label{fecnf}
f_\e^{ECNF} \cong \frac{\dD^2L_{d}}{2B^2 R_g^2 \tilde r^3c}\ 
	f^{syn}_{\bar{\e}_{syn}}
\end{equation}
where
\begin{equation}
\label{eecnf}
\bar{\e}_{syn} = \bar{\e}_{syn}(\tilde r) = {2\e\e_B\over \dD \bar\e(\sqrt{3 r})}
\cong 3 \times 10^{-10}\ \frac{\e B({\rm G})}{\dD}\ 
	\left(\frac{M_8\eta \tilde r^3}{l_{Edd}}\right)^{1/4}\ .
\end{equation}
In the far field ($\tilde r\gg\Gamma^4$), 
\begin{equation}
\label{fecff}
f_\e^{ECFF} \cong \frac{3}{8}\ \frac{L_{d}}{r^2cB^2\dD^2}\ 
	f_{\bar{\bar{\e}}_{syn}}\;,
\end{equation}
where
\begin{equation}
\label{eecnf1}
\bar{\bar{\e}}_{syn} = 3.2\times10^{14}\ \dD^4\e\e_BM_8\ 
	\left(\frac{\eta}{l_{Edd}}\right)^{1/4}
\end{equation}
\citep{ds02}.

\subsubsection{$\g\g$ Opacity from Accretion Disk Photons}

Photons from the accretion disk will interact with high energy 
$\g$-rays to produce electron-positron pairs, modifying the very-high energy
(VHE; multi-GeV -- TeV) 
$\g$-ray spectrum by a factor of $e^{-\tau_{\g\g}}$.  
The absorption opacity $\tau_{\g\g}$ can be calculated 
by inserting the 
photon density, $n_{ph}$ for an accretion disk 
into eq. (\ref{dtauggdx}) and integrating $x$ from $r$ to $\infty$.
For a Shakura-Sunyaev accretion disk, the photon density is given by 
\begin{equation}
n_{ph}^{SS}(\e_*,\Omega_*) = \frac{I_{\e_*}^{SS}(\Omega_*; \tilde{R})}{\e_* m_ec^3}
\end{equation}
where $I_\e^{SS}(\Omega; \tilde{R})$ is given by eq. (\ref{IeSS}).
For an azimuthally symmetric accretion disk, 
the optical depth to $\gamma\gamma$ pair production attenuation 
for a photon with observed energy $\e_1$ traveling outward along the jet axis
starting at height $\tilde r$ 
is given by
\begin{equation}
\tau_{\g\g}^{SS}(\e_{1},\tilde r) \cong  3 \times 10^{6}\  
	\frac{l_{Edd}^{3/4}M_8^{1/4}}{\eta^{3/4}}\ 
	\int^{\infty}_{\tilde{r}} \frac{d\tilde{x}}{\tilde{x}^2}\ 
 \int^{\infty}_6 {d\tilde{R}\over 
	\tilde{R}^{5/4}}\;
{[\phi(\tilde R )]^{1/4}H\left(\tilde s - 1\right)\over (1+\tilde R^2/\tilde x^2)^{3/2}}\;  
	\left[{\sigma_{\g\g}(\tilde s)\over \pi r_e^2}\right]\; (1-\mu_*)	\; ,
\label{tauggSStilder}
\end{equation}
where $\tilde s \equiv \langle\e(\tilde{R})\rangle {\e_1(1+z)(1-\mu_*)/2}$ and 
$
\mu_* = 1/\sqrt{1+\tilde{R}^2/\tilde{x}^2}.
$
For the Shakura-Sunyaev disk extending to the 
innermost stable orbit of a Schwarzschild black hole,  
one sees from eq.\ (\ref{tauggSStilder}) 
and the definition of $\langle\e(\tilde{R})\rangle$, eq.\ (\ref{<e>_1}), and $\xi$, 
eq.\ (\ref{xidef1}),  
 that
\begin{equation}
{\tau_{\g\g}^{SS}(\e_{1},\tilde r)\over M_8} \propto {\rm ~function~of~}\xi\;{\rm and~}\tilde r \; .
\label{xidef}
\end{equation}
A first-order correction to $\gamma\gamma$ opacity for a photon 
traveling at a small angle angle $\theta_s =\arccos \mu_s \ll 1$ 
along the jet axis is obtained by replacing $\mu_*$ with $\mu_*\mu_s$, 
which implicitly assumes that typical interactions take place at azimuth 
$\phi_* = \pi/2$. 
A detailed calculation of the $\g\g$ opacity from accretion disk photons is given by \citet{bk95}.

\subsubsection{Numerical Results}

\begin{deluxetable}{lll}
%\rotate
\tabletypesize{\scriptsize}
\tablecaption{
Parameters of Baseline Model
}
\tablewidth{0pt}
\startdata
\hline
Redshift & 	$z$		& 1	  \\
Bulk Lorentz Factor & $\Gamma$	& 25	  \\
Doppler Factor & $\dD$       & 25    \\
Magnetic Field & $B$         & 1 G   \\
Variability Timescale & $t_v$       & $10^4$ s \\
Black Hole Mass ($10^8 M_\odot$) & $M_8$       & 1   \\
Eddington Ratio & $l_{Edd}$   & 1     \\
Jet Height (units of $R_g$) & $\tilde{r}$ & $10^3$ \\
Low-Energy Electron Spectral Index & $p_1$       & 2     \\
High-Energy Electron Spectral Index  & $p_2$       & 4	 \\
Minimum Electron Lorentz Factor & $\gp_{min}$  & $10^2$ \\
Maximum Electron Lorentz Factor & $\gp_{max}$  & $10^7$ \\
Accretion Efficiency & $\eta$ & 1/12 \\
Ratio of $B$ to Equipartition Field & $B/B_{eq}$ & 0.3 \\
Jet Power in Magnetic Field & $P_{j,B}$ & $7\times10^{43}$ erg s$^{-1}$ \\
Jet Power in Particles & $P_{j,par}$ & $10^{46}$ erg s$^{-1}$ \\
\enddata
\end{deluxetable}

The electron distribution is assumed to be well-represented by 
the \citet{band93}-type function
$$N^\prime_e(\gp) = K^\prime_e\ H(\gp;\gp_{min},\gp_{max})\;\;\big\{
	 \g^{\prime -p_1} \exp\left(-\gp/\gp_0\right)\;H[(p_2-p_1)\gp_0 - \gp] \;
	+$$
\begin{equation}
[(p_2-p_1)\gp_0]^{p_2-p_1} \g^{\prime -p_2}\ \exp(p_2-p_1) H[\gp -(p_2-p_1) \gp_0]
	\big\}.
\label{bandfn}
\end{equation}

This distribution is essentially two smoothly joined power laws with
power law number indices $p_1$ and $p_2$, and low- and high-energy
cutoffs, $\gp_{min}$ and $\gp_{max}$, respectively, in the electron
spectrum.  For illustrative purposes, we take $p_1 = 2$ and $p_2 = 4$,
or a break by 2 units in the electron spectrum. This can be compared
to a break by one unit expected for synchrotron and Thomson
losses. Our approach is, however, to use the flaring synchrotron
spectrum to imply the underlying flaring electron distribution without
regard to specific acceleration and radiation processes. When the
nonthermal electron distribution is obtained from analysis of blazar
data, then the underlying jet physics that gives rise to the inferred
electron spectrum can be examined.

The total jet power in the stationary frame of the host galaxy 
is given by 
\begin{equation}
\label{jetpower}
P_j = 2 \pi R_b^{\prime 2}\beta\G^2c u^{\prime}_{tot}
 = P_{j,par}+P_{j,B}
\end{equation}
\citep{celotti93,celotti07,fdb08}, where $u^{\prime}_{tot}$ is the
total energy density in the jet, $P_{j,par}$ is the jet power from
particles, and $P_{j,B}$ is the jet power from the magnetic field.
Here the factor of $2$ takes into account that the jet is two-sided.
For synchrotron--only emission, it is expected that the jet will be in
equipartition and $P_{j,par} \approx P_{j,B}$, which will minimize the
jet power.  In order to explain the Compton-scattered component,
however, the energy density in electrons can be different than the
magnetic field energy density.

We performed a parameter study by varying model parameters with
respect to a baseline model, with baseline parameters given in Table
1.  We consider a $10^8 M_\odot$ supermassive black-hole jet source
located at redshift $z = 1$. The jet is radiating at $10^3 R_g$ from
the black hole and has Doppler factor $\dD = 25$ and bulk Lorentz
factor $\Gamma = 25$, so that the angle of the jet direction to the
line of sight is $\theta \cong 1/\Gamma$.  The mean magnetic field is
1 G, and the variability time $t_v = 10^4$ s, corresponding to
$\approx 10\times$ the light crossing time for the Schwarzschild
radius of a $10^8 M_\odot$ black hole. The jet opening angle $\theta_j
\approx R_b^\prime /r \lesssim c\dD t_v/(1+z) r \cong 0.25 \approx
15^\circ$ for standard parameters.  While varying the parameters, the
synchrotron spectrum was kept relatively constant by varying
$K^\prime_e$ and $\gp_0$ following the relations given in \S
\ref{trends} (except when changing angle).  This was done with a
$\chi^2$ fitting technique to the baseline synchrotron spectrum
\citep[see][]{fdb08}.  The transition between the near field and far
field takes place at $\tilde r \approx \Gamma^4$, so that the baseline
height of the jet is in the near field.

%\clearpage

\begin{figure}[t]
\center
{\includegraphics[scale=0.5]{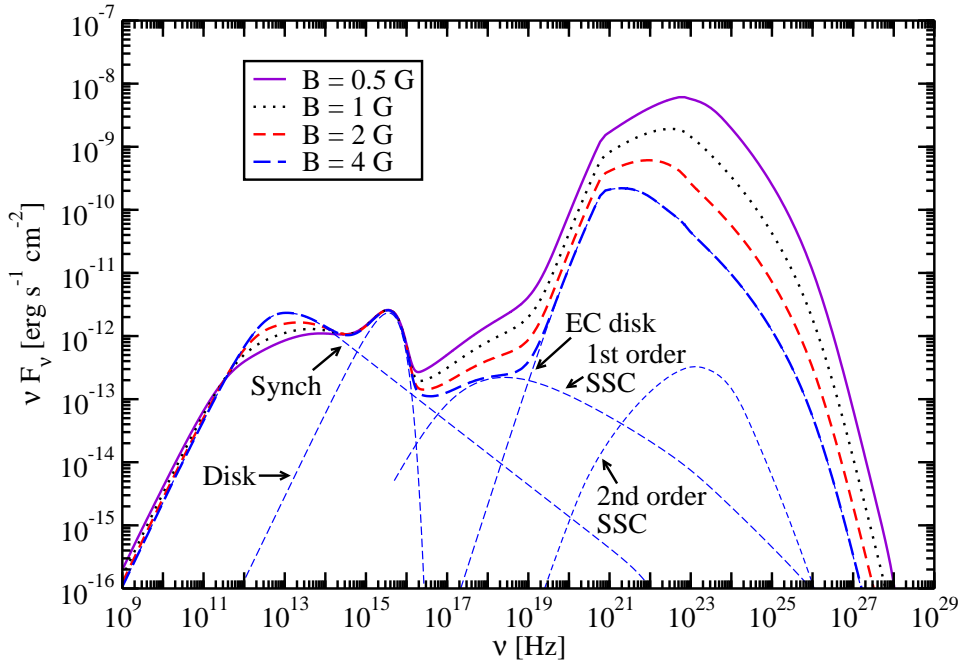}}
\caption[]{Spectral energy distribution (SED) of
accretion-disk/relativistic jet model using standard parameters for
disk-jet system given by Table 1. Magnetic field is varied with the
synchrotron flux remaining essentially constant, though with the peak
synchrotron frequency varying between $\approx 2\times 10^{11}$ and
$10^{13}$ eV. Separate spectral components (thin short dashed
curves) are, from left to right, the synchrotron, accretion-disk,
SSC, Compton-scattered accretion disk radiation, and second-order
SSC components, for the high-magnetic field case, $B = 4$ G.
}
\label{f1}
\end{figure} 

%\clearpage

The Compton-scattered accretion disk spectra are calculated from eq.\
(\ref{eslcssexyz}).  Fig.\ \ref{f1} shows the effects of changing the
magnetic field. With increasing $B$, fewer electrons are required to
make the same synchrotron flux, so that both the first and 
second-order SSC flux, and the flux of the Compton-scattered accretion-disk
component decrease with increasing $B$.  In all of these models, the
second-order SSC emission is overwelmed by the external Compton
component and is not visible.  The overall levels of all
Compton-scattered components are $\propto B^{-2}$.

%\clearpage

\begin{figure}[t]
\center
{\includegraphics[scale=0.5]{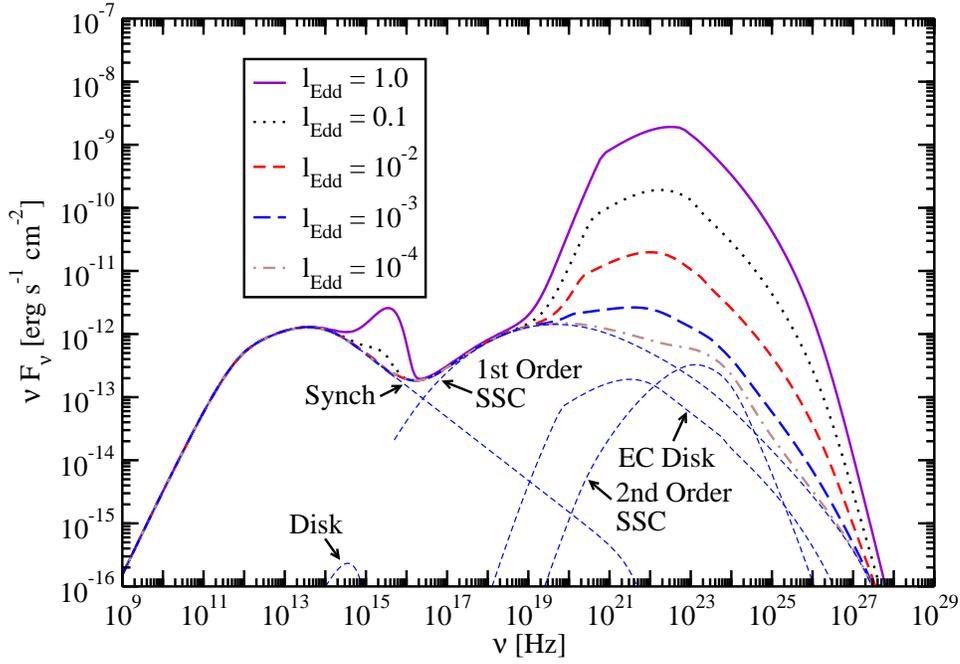}}
\caption[]{ Same as Fig.\ \ref{f1}, except that the luminosity of the
disk is changed.  Separate spectral components are shown for the
$l_{Edd}=10^{-4}$ case.}  
\label{f2}
\end{figure} 

%\clearpage

\begin{figure}[t]
\center
{\includegraphics[scale=0.5]{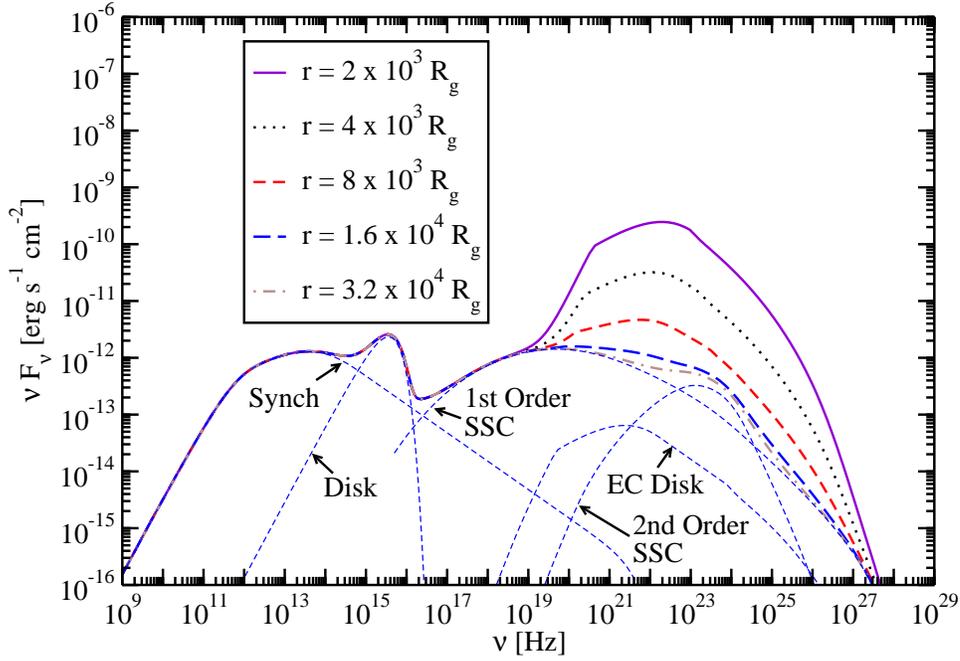}}
\caption[]{Same as Fig.\ \ref{f1}, except that the jet height is varied.
Separate spectral components are shown for the
$r=3.2\times10^{4}\ R_g$  case.}   
\label{f3}
\end{figure} 

%\clearpage

\begin{figure}[t]
\center
{\includegraphics[scale=0.5]{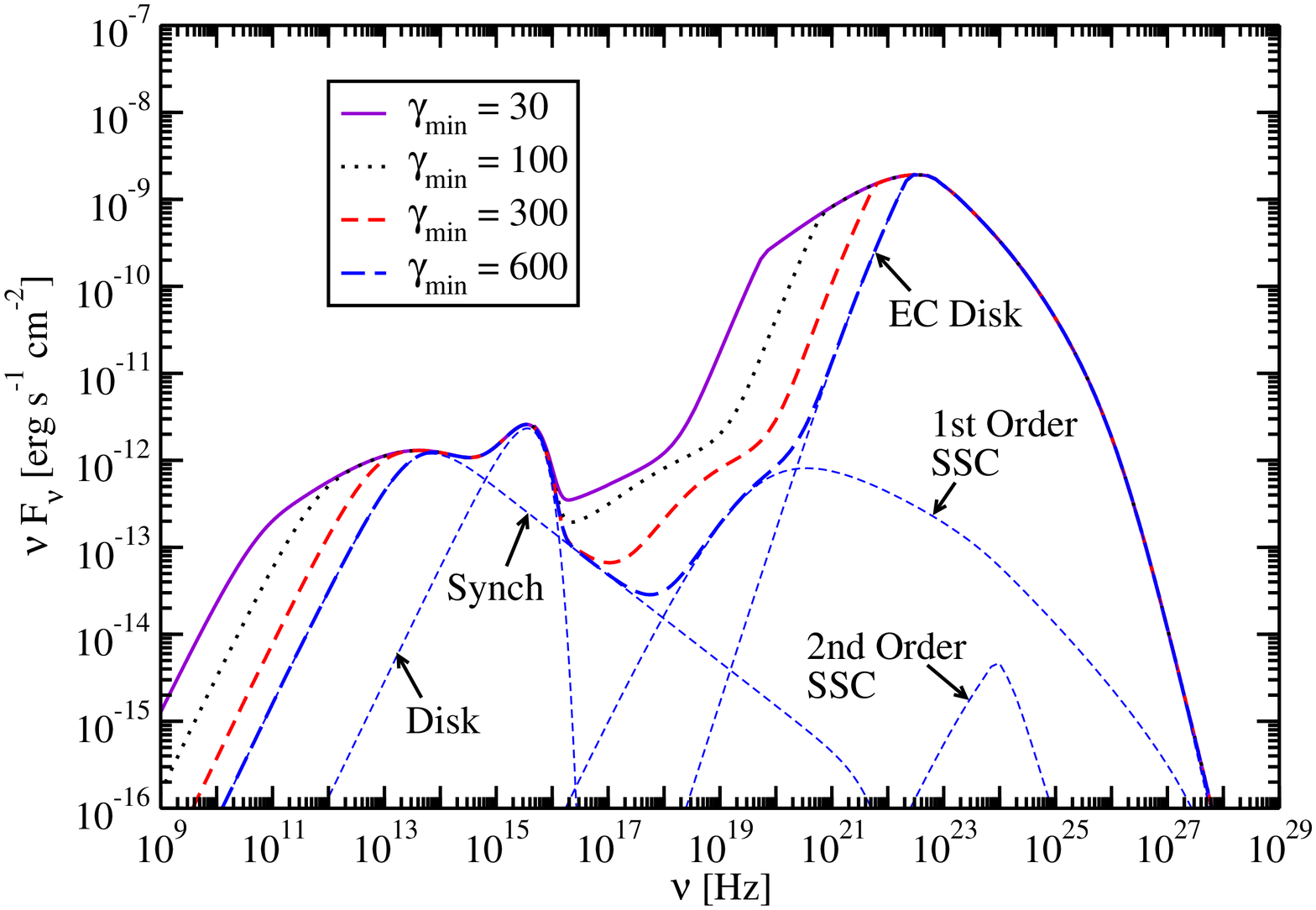}}
\caption[]{
Same as Fig.\ \ref{f1}, except that the minimum electron Lorentz factor $\gp_{min}$ is changed.  
Separate spectral components are shown for the
$\g_{min}=600$ case.}  
\label{f4}
\end{figure} 

%\clearpage

With increasing disk power $\ell_{\rm Edd}$, the accretion-disk
radiation and also the Compton-scattered accretion-disk
component becomes progressively larger, as shown in Fig.\ \ref{f2}.
Note that the temperature of the disk photons at a given radius
increases $\propto \ell_{\rm Edd}^{1/4}$.  Fig.\ \ref{f3} displays the
dependence of the blazar SED on height $\tilde r$ of the jet. As the
blob's distance from the disk increases, the level of the Comptonized
disk radiation decreases.  In both the case of changing the
$\ell_{\rm Edd}$ and $\tilde r$ the SSC components are unaffected;
eventually, the Compton-scattered disk flux falls below the SSC fluxes.
The second-order SSC is visible when the external Compton flux
decreases enough.  Fig.\ \ref{f4} shows how variations in $\g_{min}$
affect the low energy part of the synchrotron, SSC, and external
Compton components.

%%%%%%%%%%%%%%%%%%%%%%%%%%%%%%%%%%%

%\clearpage

\begin{figure}[t]
\center
{\includegraphics[scale=0.5]{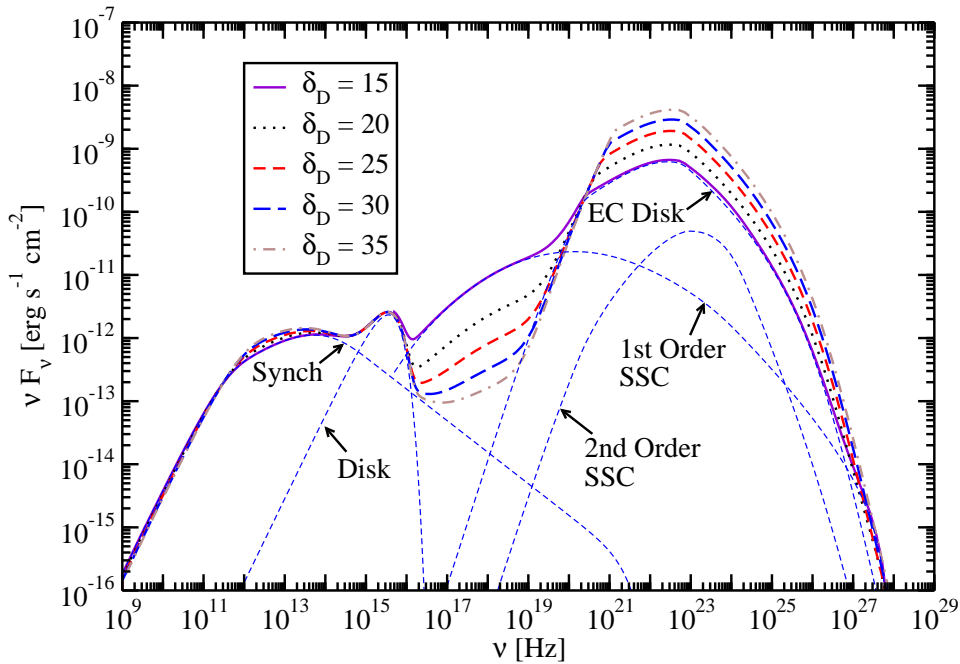}}
\caption[]{
Same as Fig.\ \ref{f1}, except that the Doppler factor is varied with the synchrotron
component essentially held constant.  
Separate spectral components are shown for the
$\dD=15$ case.}  
\label{f5}
\end{figure} 

%\clearpage

In Fig.\ \ref{f5}, the Doppler factor is varied from the baseline
value of 25, with the same approximate synchrotron flux.  As $\dD$
increases, the SSC component decreases, while the Compton-scattered
accretion disk component increases. This behavior follows from
eqs. (\ref{fssc}) and (\ref{fecnf}), for the following reasons: For
larger Doppler factors and fixed variability times, the radius becomes
larger and fewer electrons are required to make the synchrotron flux
at the same flux level, so the electron and internal photon energy
densities decrease. Consequently the SSC flux decreases, but the
larger value of $\dD$ means that the external target photon field
becomes more intense. Thus the Compton-scattered accretion disk
radiation increases with increasing $\dD$.  Also, as the Doppler
factor increases, the SSC component is shifted to lower energies while
the Compton-disk component is shifted to higher energies, in
accordance with eqs. (\ref{essc}) and (\ref{eecnf}), respectively.

%\clearpage

\begin{figure}[t]
\center
{\includegraphics[scale=0.5]{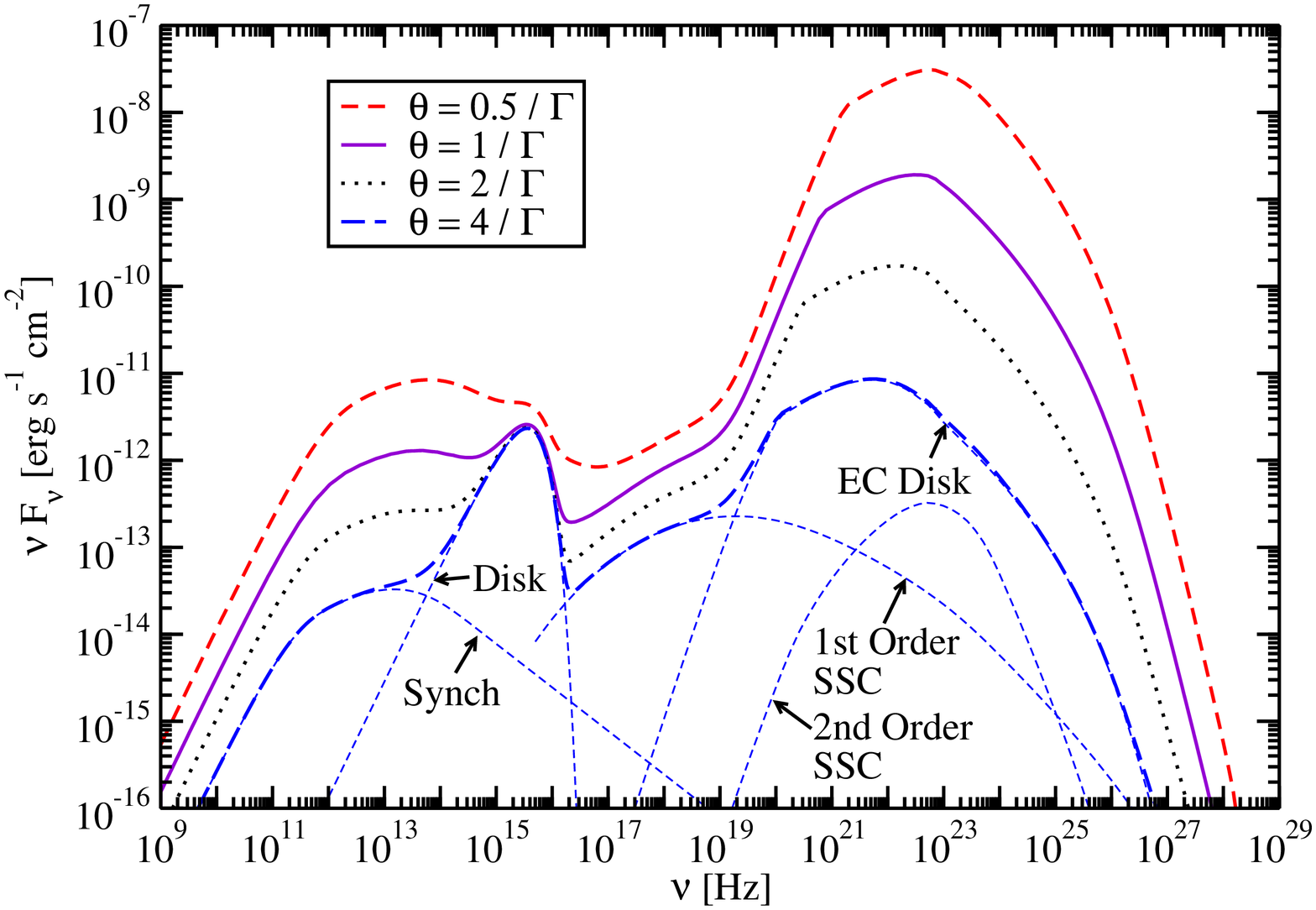}}
\caption[]{Same as Fig.\ \ref{f1}, except that the viewing angle is varied. The peak flux of 
the synchrotron, SSC, and near-field scattered disk component goes roughly 
as $\dD^4$, $\dD^2$, and $\dD^6$, 
respectively. The different observing angles $\theta = 1/\Gamma, 2/\Gamma,3/\Gamma, 4/\Gamma$
correspond to Doppler factors $\dD = 25, 10, 5,$ and 2.94, respectively. 
Separate spectral components are shown for the
$\theta = 4/\Gamma$ case.}  
\label{f6}
\end{figure} 

%\clearpage

Fig.\ \ref{f6} illustrates how the blazar SED is affected by changes
in viewing angle. Contrary to Figs.\ (\ref{f1}) -- (\ref{f5}), we let
the synchrotron flux vary. In this calculation, $\Gamma = 25$, and the
observation angle and therefore $\dD$ changes. It is interesting to
note that the SSC flux varies least with changes in viewing angle,
while the Comptonized disk component changes most rapidly.  The ratios
of the various components are explained by noting that $\dD/\Gamma
\cong 1, 2/5, 1/5,$ and 2/17 for $\theta \Gamma = 1, 2, 3,$ and 4,
respectively.  The beaming factor for synchrotron radiation is
$\propto \dD^{3+\alpha} \propto \dD^4$ in the flat portion of the $\nu
F_\nu$ SED, with the convention that flux density $F_\nu \propto
\nu^{-\alpha}$. The relative magnitudes of the SSC components are
explained from eq.\ (\ref{fssc}), noting that
$f_{\e_{pk}^{syn}}^{syn}\propto \dD^4$, so that
$f_{\e_{pk}^{SSC}}^{SSC} \propto \dD^2$.  The peak flux of the
scattered disk component in the near-field regime, $f_\e^{ECNF}
\propto \dD^6$, from eq.\ (\ref{fecnf}).

A larger viewing angle or smaller Doppler factor implies a smaller
size scale of the emitting region for a given variability time $t_v$.
The characteristic Thomson scattering depth is $n^\prime_e \sigma_{\rm
T} R^\prime_b\propto R^{\prime~-2}_b$ for a constant comoving electron
spectrum $N^\prime_e(\gp )$ (described in our calculations by eq.\
[\ref{bandfn}]) in a region of size $R^\prime_b$. Consequently, the
suppression of the SSC component due to Doppler boosting is offset by
the increased scattering depth for larger observing angles.

%\clearpage

\begin{figure}[t]
\center
{\includegraphics[scale=0.5]{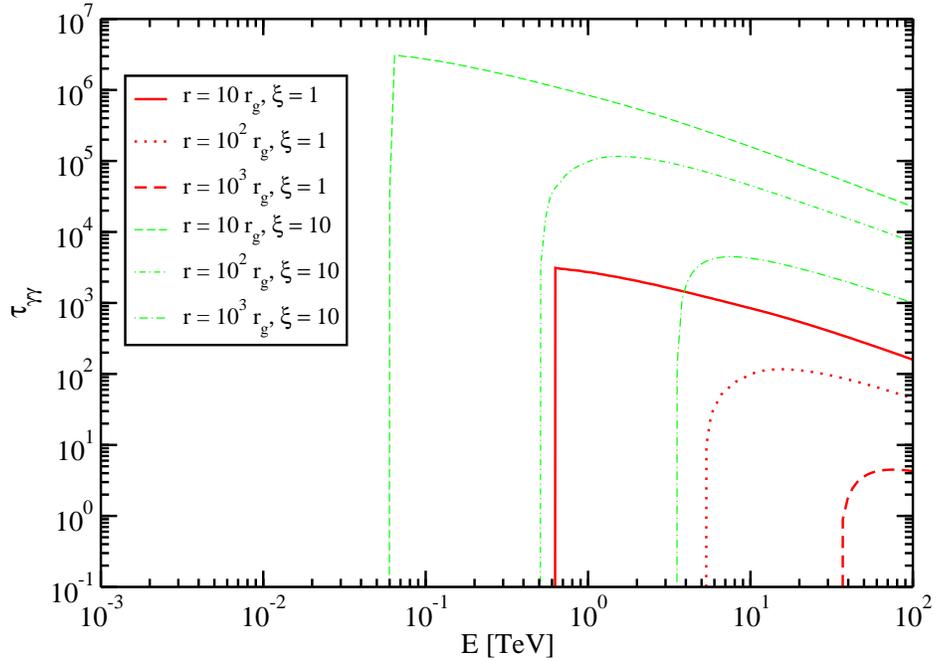}}
\caption[]{Opacity of a photon emitted at 
various jet heights $\tilde r$ in units of 
$R_g$, subject to $\gamma\gamma$ pair production
attenuation with photons from a Shakura-Sunyaev
accretion disk with parameters given in Table 1. The parameter
$\xi$ is defined in eq.\ (\ref{xidef}).}  
\label{tau_disk}
\end{figure} 

%\clearpage

Fig.\ \ref{tau_disk} shows a calculation of the accretion-disk opacity
$\tau_{\g\g}(\e_1,\tilde r)$ for photons traveling along the jet axis,
using eq.\ (\ref{tauggSStilder}) and the parameters given in Table 1.
Here we assume that the Shakura-Sunyaev disk reaches to the innermost
stable orbit ($\tilde R_{i}=6$) of a Schwarzschild black hole. At
larger jet heights, the accretion-disk opacity declines and only
increasingly higher energy photons are subject to attenuation due to
the threshold condition.  The effect of increasing $\xi$ is to
increase the mean accretion-disk photon energy at a given radius, from
eq.\ (\ref{xidef1}), and therefore lower the energy at which $\gamma$
rays can be attenuated.  At $\tilde r = 10^3$, only $\gtrsim 10$ TeV
photons are attenuated by the disk radiation field (with $\ell_{\rm
Edd} = 1$). The $\tau_{\g\g}$ corrections have not been included in
the SED calculations, but are only important at $\nu \gtrsim 10^{25}$
Hz.

In all of our models the jet is particle-dominated ($P_{j,par}\gg
P_{j,B}$) and the magnetic field is below its equipartition value. The
total jet power, also divided into magnetic field and
particle powers, is shown in Table \ref{jetpowertable}. This is typical of
results from modeling the observed X-ray and $\g$-ray emission of
TeV blazars with the SSC component \citep{fdb08}.

\begin{deluxetable}{llllll}
\tabletypesize{\scriptsize}
\tablecaption{Jet Powers for Models in Figs.\ \ref{f1} -- \ref{f6}.\tablenotemark{a}}
\tablewidth{0pt}
\tablehead{
\colhead{ $B$ [G] } &
\colhead{ $\delta_D$ } &
\colhead{ $\gamma_{min} $ } &
\colhead{ $L_B$ [$10^{43}$ erg s$^{-1}$] } &
\colhead{ $L_{par}$ [$10^{45}$ erg s$^{-1}$] } &
\colhead{ $L_{tot}$ [$10^{45}$ erg s$^{-1}$] } 
}
\startdata
1 &     25 &    100 &   6.6 & 11  & 11  \\
1 &     15 &    100 &   .85 & 36  & 36  \\
1 &     20 &    100 &   2.7 & 19  & 19  \\
1 &     30 &    100 &   14  & 7.5 & 7.6 \\
1 &     35 &    100 &   25  & 5.3 & 5.5 \\
0.5 &   25 &    100 &   1.6 & 27  & 27  \\
2 &     25 &    100 &   26  & 4.8 & 5.0 \\
4 &     25 &    100 &   110 & 2.0 & 3.1 \\
1 &     25 &    30  &   6.6 & 18  & 18  \\
1 &     25 &    300 &   6.6 & 5.7 & 5.7 \\
1 &     25 &    600 &   6.6 & 3.0 & 3.0 \\
\enddata
\label{jetpowertable}
\tablenotetext{a}{First model is baseline model.}
\end{deluxetable}

\subsection{External Isotropic Radiation Field}

This is the case treated by \citet{gkm01}, where jet electrons
scatter a surrounding external isotropic radiation field
$u_*(\e_*,\Omega_* ) = u_*(\e_* )/4\pi$.  By integrating
eq.\ (\ref{esLCesomegas}) over the angle variables, recognizing that
$d\mu_* d\phi* = 2 \pi d\cos\psi$ for this geometry, one obtains 
\begin{equation}
f^{{\rm C},iso}_{\e} = {3\over 4}\;{c\sigma_{\rm T} \e_s^2\over 4\pi d_L^2}\;\dD^3
\;
\int_0^\infty d\e_*\;{u_*(\e_* )\over\e_*^2}
\int_{\gamma_{min}}^{\gamma_{max}} d\gamma\;
{N^\prime_e(\g/\dD )\over \gamma^2 } \; F_{\rm C}(q,\Gamma_e)\;,
\label{jesCpowerlaw}
\end{equation}
where $F_{\rm C}(q,\Gamma_e)$ is given by Jones's formula  \citep{jon68},
eq.\ (\ref{fcq}). 
Because the scattering is taking place in the stationary frame, 
\begin{equation}
q \equiv {\e_s/\g \over \Gamma_e
(1-\e_s/\g )}\;,
\label{qnormal}
\end{equation}
with $\Gamma_e = 4\e_*\g\;$ and, as before, $ \e_s = (1+z) \e$.
The limits $\gamma_{min}$
and $\g_{max}$ are given by 
\begin{equation}
\g_{min} = {1\over 2} \e_s\;\left( 1+\sqrt{1+{1\over \e_*\e_s}} \;\right)
\label{gpmin1}
\end{equation}
and
\begin{equation}
\g_{max} = {\e_*\e_s\over \e_* - \e_s}H(\e_* -  \e_s) \;+\; \g_2H(\e_s - \e_* )\;.
\label{gpmax1}
\end{equation}

For the
case of the cosmic microwave background radiation (CMBR),
\begin{equation}
{u_*(\e_*)\over 4\pi }\;= \; u_{bb}(\e_*,\Omega_* ) = {2m_e c^3\over \lambda_{\rm C}^3}\;
{\e_*^3\over \exp(\e_*/\Theta ) - 1 }\;,
\label{blackbody}
\end{equation}
where $\Theta = \kB T/m_ec^2$ is the dimensionless temperature
of the blackbody radiation field, and $T = 2.72(1+z)$ K.  
Substituting eq.\ (\ref{blackbody})
in eqs.\  (\ref{fecompton}) and (\ref{esLCesomegas_1}) gives
\begin{equation}
f_\e^{{\rm C},CMBR} = {3 m_ec^3\sigma_{\rm T}\e_s^2 \dD^3\over 2 d_L^2 \lambda_{\rm C}^3}
\int_0^\infty d\e_*\; {\e_*\over \exp(\e_*/\Theta ) - 1 }\int_{\g_{min}}^{\g_{max}} d\gamma
\; \gamma^{-2} N^\prime_e(\g/\dD  ) \;F_{\rm C} (q,\Gamma_e)\;. 
\label{eslcsses}
\end{equation}
Eq.\ (\ref{eslcsses}) gives an accurate
spectrum of radiation made when jet electrons with number spectrum
$N^\prime_e(\gp )$ Compton-scatter blackbody
photons \citep[see, e.g.][]{tav00,da02,bdf08}.

\subsection{Scattered BLR Radiation Field}

The BLR is thought to consist of dense clouds with a specified covering factor 
that can be determined from AGN studies. These clouds
intercept central-source radiation
to produce the broad emission lines in broad-line AGN 
\citep{kn99}. The diffuse gas will also 
Thomson scatter the central source radiation. The scattered radiation provides an important 
source of target photons that jet electrons scatter to $\gamma$-ray energies \citep{sbr94,sik97}.
This radiation field also attenuates $\gamma$ rays produced within
the BLR \citep{bl95,lb95,bd95,dp03,lb06,rei07,sb08}.

%\clearpage

\begin{figure}[t]
\center
{\includegraphics[scale=0.5]{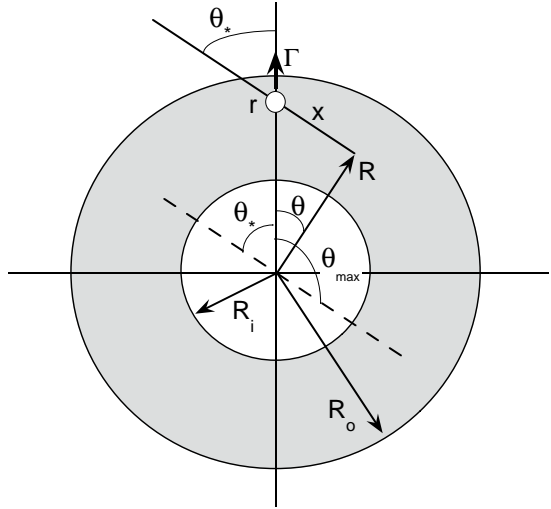}}
\caption[]{
Idealized geometry of the scattering region, approximated as a spherically-symmetric shell of 
gas with radial Thomson depth $\tau_{\rm T}$ between inner and outer radii
$R_i$ and $R_o$, respectively, and a density gradient defined by index $\zeta$, where
$n_e(R) \propto R^{\zeta}$. }  
\label{f8}
\end{figure} 

%\clearpage

Here we calculate the angular distribution of the Thomson-scattered radiation 
from a shell of gas with density 
\begin{equation}
n_e(R)\; = n_0 \left( {R\over R_i} \right)^{\zeta}\;H(R;R_i,R_o)
\label{ner}
\end{equation}
extending from inner radius $R_i$ to outer radius $R_o$ (see Fig.\ \ref{f8}; 
the calculation of fluorescence
atomic-line radiation differs by considering dense clouds with a volume filling
factor). 
The shell is assumed to be spherically symmetric with a power-law radial density
distribution and radial Thomson depth  $\tau_{\rm T} = \sT\int_{R_i}^{R_o} dR\; n_e(R)$. 
The Thomson-scattered spectral photon density can be estimated by noting
that a fraction $\approx rn_e(r)\sT $ of the central source radiation with ambient
photon density $n_{ph}(\e_*; r) = \dot N_{ph}(\e_* )/4\pi r^2 c$ is scattered, giving 
a target scattered radiation field 
\begin{equation}
n_{sc}(\e_*;r)\approx {n_e(r)\sT \dot N_{ph}(\e_*)\over 4\pi r c}\;
\label{nscer}
\end{equation}
when $R_i \lesssim r \lesssim R_0$. 
Here $\dot N_{ph}(\e_* ) = L(\e_* )/(m_ec^2 \e_* )$ is the central source 
photon-production rate, assumed to radiate isotropically,
and $L(\e_* )$ is its spectral luminosity.

A more accurate calculation of the Thomson-scattered photon density is obtained by integrating
the expression 
\begin{equation}
n_{ph}(\e_* ; R)\; = \; \int dV \; {\dot n (\e_*; \vec R ) \over 4\pi x^2 c}\;
\label{nph}
\end{equation}
over volume, where  $dV = R^2 d\phi d\mu dR$ and $x^2 = R^2 + r^2 -2rR\cos\theta$ (Fig.\ \ref{f8}).
Assuming that the photons
are isotropically Thomson-scattered by an electron
 without change in energy, so that $ \dot n (\e_*; \vec R )
= \dot N_{ph}(\e_*)\sT n_e(R)/(4\pi R^2 c)$, then
\begin{equation}
n_{ph}(\e_* ; r)\; 
= \;
{L(\e_* )\sT \over 8\pi m_ec^3 \e_*} \int_{-1}^1 d\mu\; \int_{0}^\infty dR \; {n_e(R)\over x^2}
\;= \;
{L(\e_* )\sT \over 8\pi m_ec^3 \e_* r} \int_{0}^\infty dR \; {n_e(R)\over R}\;
\ln\left|{R+r\over R-r}\right|
\label{nph7}
\end{equation}
 from a 
spherically-symmetric electron density distribution
\citep[cf.][for a time-dependent treatment]{gou79,bd95}. In the case of an isotropic, uniform surrounding medium, 
$n_e(R) = n_0$, $\zeta = 0$, $R_i = 0$ and 
$R_o \rightarrow \infty$ in eq.\ (\ref{ner}), and eq.\ (\ref{nph7}) gives 
$n_{ph}(\e_*; r) = 3.324\dot N_{ph}(\e_*)\sT n_0/ (8\pi r c)$, noting that the integral $\int_0^\infty du\;
\ln|(1+u)/(1-u)|/u \cong 3.324$. 
Thus the approximation given by eq.\ (\ref{nscer}) is accurate to 
within a factor of $\approx 2$ for this case.

The angle-dependent scattered photon distribution $n_{ph}(\e_*, \mu_*; r)$ can 
be derived by imposing a $\delta$-function constraint on the angle 
$\theta_* = \arccos \mu_*$ in eq.\ (\ref{nph}) \citep[see also][]{dp03}, so that
\begin{equation}
n_{ph}(\e_*, \mu_*; r)\;=\;{\sT \dot N_{ph}(\e_* )\over 8\pi c r}\;\int _{-\mu_*}^1 d\mu\;\int_0^\infty dg\;
{n_e(gr)\delta [\mu_* - \bar\mu_*(\mu,g)]\over g^2 + 1 - 2g\mu }\;,
\label{nphemuz}
\end{equation}
after changing variables to $g = R/r$. From Fig.\ \ref{f8}, $\theta_{max} = \pi - \theta_*$, 
so that $\mu_{min} = \cos\theta_{max} = \pi - \theta_*$ and $-\mu_* \leq \mu \leq 1$.
The law of sines gives $R^2 (1-\mu^2) = x^2 (1-\mu_*^2)$, with the result
\begin{equation}
\bar\mu_*(\mu,g) \;=\; {\pm (1-g\mu)\over \sqrt{1+g^2 - 2 g \mu}}\;.
\label{barmustarmuy}
\end{equation}
Transforming the $\delta$-function in $\mu_*$ to a $\delta$-function in $g$ 
gives, after solving, 
\begin{equation}
n_{ph}(\e_*,\mu_*;r) \; = \; {\sT \dot N_{ph}(\e_* )\over 8\pi c r }\;
{\cal N}(\mu_*,r)\;,
\label{nphemuz1}
\end{equation}
where 
\begin{equation}
{\cal N}(\mu_*,r) \;=\; {\cal N}[\mu_*,n(r)]\;\equiv \;
\int_{-\mu_*}^1 d\mu\; n_e(\bar gr) \;{\sqrt{1+\bar g^2 - 2\bar g \mu} \over \bar g ( 1- \mu^2) }\;
\label{nphemuz111}
\end{equation}
 (units of ${\cal N}$ are $1/L^3$), and
\begin{equation}
\bar g \;=\; \bar g(\mu,\mu_*) \;\equiv\; {-\mu (1- \mu_*^2) + \mu_*\sqrt{(1-\mu^2 )(1-\mu_*^2)}\over 
\mu_*^2 -\mu^2 }\;.
\label{gbar}
\end{equation}

%\clearpage

\begin{figure}[t]
\center
{\includegraphics[scale=0.6]{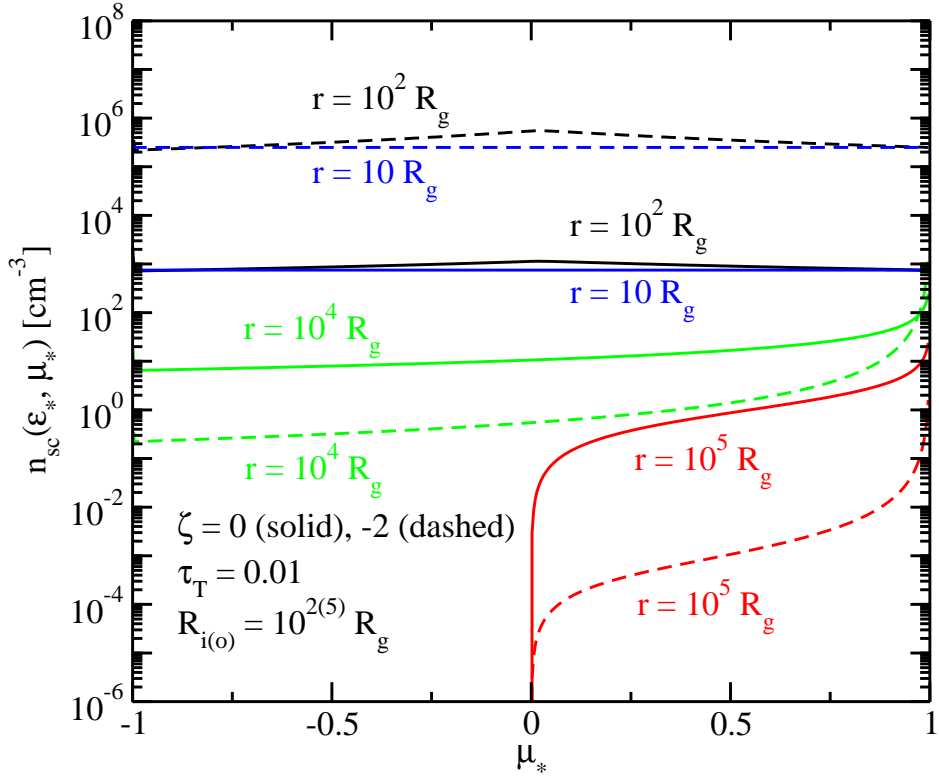}}
\caption[]{
Angle-dependent density of Thomson-scattered radiation at height $r$ along
the jet axis. The radial Thomson depth of the spherically-symmetric scattering
medium is $\tau_{\rm T} = 0.01$, and the scattering 
shell is assumed to extend from $10^2 $ to $10^5$ Schwarzschild 
radii for a $10^8 M_\odot$ black hole. The solid and dashed curves 
show results with $\zeta = 0$ and $\zeta = -2$, respectively. The values of $r$ for
are labeled on the dashed curves. The ratio of the luminosity of the photon
source to the dimensionless photon energy is $L_*/\e_* =10^{44}$ ergs s$^{-1}$, and 
the calculation assumes that $\e_*\ll 1$, 
so that scattering is in the Thomson regime (see eq.\ [\ref{dotNphestar}]).}  
\label{fignsc}
\end{figure} 

%\clearpage

Fig.\ \ref{fignsc} shows the angle dependence of the scattered radiation field
in the stationary frame for this idealized geometry for the parameters 
given in the figure caption. As can be seen, the scattered radiation field 
is nearly isotropic when $r < R_i$, 
and starts to display increasing asymmetry peaked in the outward direction at
increasingly greater heights. When $r > R_o$, all scattered radiation is outwardly 
directed. Most of the scattering material is near the inner 
edge when $\zeta = -2$, so that the intensity of the scattered radiation field 
is largest at $r\lesssim R_i$. By contrast, the intensity of the scattered radiation 
field is not so enhanced towards the inner regions when $\zeta = 0$.

The Compton-scattered radiation spectrum is given, in general, by eq.\ (\ref{esLCesomegas_26}). 
Substituting eq.\ (\ref{nphemuz1}) for the angular distribution of the
target photon source gives, for a monochromatic photon source
\begin{equation}
\dot N_{ph}(\e_* ) = {L_0\delta(\e_* - \e_{*0})\over m_ec^2 \e_*}\;,
\label{dotNphestar}
\end{equation}
the $\nu F_\nu$ flux
\begin{equation}
f_\e^{EC,scat}(r) \; = \;{(\pi r_e^2)^2 L_0 \dD^3\over 12\pi^2 d_L^2 r}\;
\left({\e_s\over \e_*}\right)^2\;\int_{\max(-1,1-2\e_s/\e_*)}^1 d\mu_*\;
{\cal N}(\mu_*,r)\;\int_{\tilde \g_{low}}^\infty d\g\;{N_e^\prime(\g/\dD )
\over \g^2}\;\Xi\;.
\label{feEC_scat}
\end{equation}
In this expression, 
\begin{equation}
\tilde \g_{low} \;\equiv \; {\e_s \over 2}\;\left[ 1 + \sqrt{1 + {2\over \e_*\e_s(1-\mu_*)}}\;\right]\;
\label{tildeglow}
\end{equation}
(compare eq.\ [\ref{glow}]).
Substituting eq.\ (\ref{nphemuz1}) into eq.\ (\ref{dtauggdx}) 
for a monochromatic photon source, eq.\ (\ref{dotNphestar}), 
gives 
\begin{equation}
\tau_{\g\g}(\e_1,r) \; = \; 
{\sigma_{\rm T} L_0 \over 8\pi m_ec^3 \e_*}\;\int_r^\infty {dx\over x}\;
\int_{-1}^{1 - 2/[\e_1\e_*(1+z)]} d\mu_*\;(1-\mu_*)\;{\cal N}(\mu_*,x)\sigma_{\g\g}(s)\;
\label{taugge1r}
\end{equation}
for the opacity of a photon with measured energy $\e_1$ emitted outward 
along the jet axis at height $r$. The $\g\g$ opacity vanishes
when $\e_1 \leq 1/[\e_*(1+z)]$ due to the $\g\g$ pair-production
threshold. 

\subsubsection{Thomson-Scattered Isotropic Monochromatic Radiation Field}

Substituting
$u_*(\e_*,\Omega_*) = u_{*0}\delta(\e_* - \e_{*0})/4\pi$
 for an isotropic monochromatic radiation field in eq.\ (\ref{esLCesomegas_2})
gives, with $\Xi \rightarrow 2$ for the Thomson regime away from the endpoints
of the spectrum, 
\begin{equation}
f_\e^{{\rm T},iso} = {c\pi r_e^2 \dD^{3+p} u_{*0}\over 4\pi d_L^2}\;
\left({\e_s \over \e_*}\right)^2\;\int_{\max(-1,1-2\e_s/\e_{*})}^1 d\mu_*\; 
\int_{\sqrt{\e_s/2\e_*(1-\mu_*)}}^\infty d\g\; \g^{-2}N_e^\prime(\g/\dD )\;.
\label{feTiso}
\end{equation}
We consider Compton up-scattering, i.e., $\e_s >\e_*$ for the power-law electron
spectrum given by eq.\ (\ref{Neprime}). 
The Thomson approximation is only valid far from the endpoints. 
The asymptotes for the Thomson-scattered spectrum of a surrounding 
isotropic, monochromatic radiation field therefore becomes
\begin{equation}
f_\e^{{\rm T},iso} \cong {2c\pi r_e^2  u_{*0}\over 4\pi d_L^2 (p+1)}\;
K^\prime\left({\e_s \over \e_*}\right)^2\;\dD^2 \g_1^{\prime~-(p+1)}\;,\;{\rm for}\;\;
\e_* \lesssim \e_s \ll 4\e_s(\dD^2\gamma_1^\prime)^2\;,
\label{feTisocase1}
\end{equation}
and 
\begin{equation}
f_\e^{{\rm T},iso} \cong {2^{p+3}\over (p+1)(p+3)}\;{c\pi r_e^2 \dD^{3+p} u_{*0}\over 4\pi d_L^2 }\;
K^\prime\left({\e_s \over \e_*}\right)^{(3-p)/2}\;,\;{\rm for}\;\;
4\e_s(\dD^2\gamma_1^\prime)^2 \lesssim \e_s \ll 4\e_s(\dD^2\gamma_2^\prime)^2\;.
\label{feTisocase2}
\end{equation}
 Note the different beaming factors in the two asymptotes.
Eq.\ (\ref{feTisocase2}) agrees with the Thomson expression 
derived by \citet{dss97}, eq.\ (22), to within factors of order unity.

\subsubsection{Numerical Calculations}

We now present calculations of the SED of FSRQ blazars, including the
external Compton-scattering component formed by jet electrons that
scatter target photons which themselves were previously scattered by
BLR material. This EC BLR component is not found in conventional
synchrotron/SSC models of blazars, and so distinguishes standard model
blazar BL Lacs and FSRQs.  The inclusion of the $\g\g$ opacity through
the scattered radiation makes the calculation self-consistent.

%\clearpage

\begin{figure}[t] \plottwo{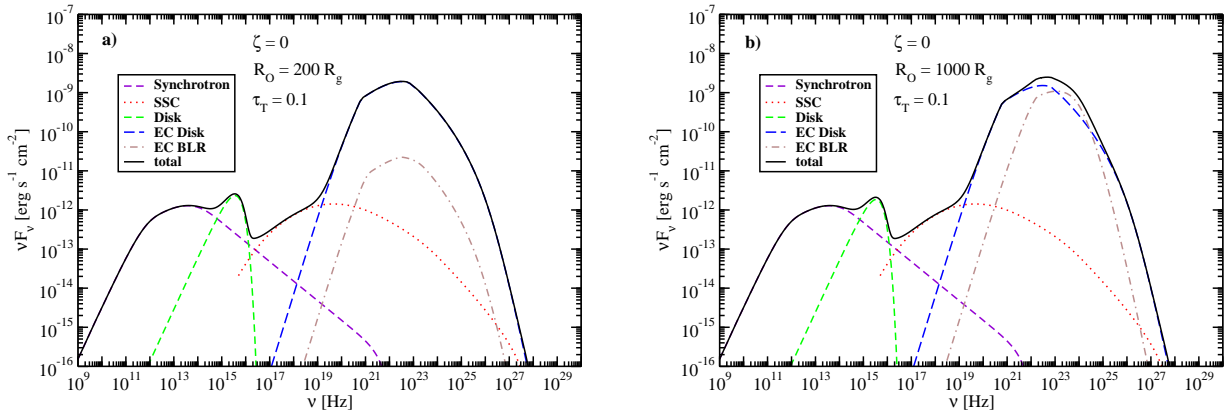}{f10b_color}
\caption{Standard  model FSRQ blazar with standard parameters
given in Table 1. The optical depth
through the BLR is $\tau_{\rm T} = 0.1$, with no 
gradient in the density of the scattering material. The radiating jet is 
located at $10^3 R_g$. Separate synchrotron, accretion-disk, SSC, EC Disk, and EC BLR
components are shown. (a) The BLR extends from $100$ to $200 R_g$.  
(b) The BLR extends from $100$ to $1000 R_g$.} 
\label{f10} 
\end{figure}

%\clearpage

For the simplified shell geometry depicted in Fig.\ \ref{f8},  
we calculate the external Compton scattering component using eq.\ (\ref{feEC_scat}), and calculate
the $\g\g$ opacity using eq.\ (\ref{taugge1r}). To simplify
the calculations, the spectrum of the radiation scattered by the BLR
is assumed to be monochromatic with energy $\approx 50$ eV,
corresponding to the mean energy from the accretion-disk radiation for 
the standard model.
Results of such a calculation are shown in Fig.\ \ref{f10}, using parameters for 
a standard FSRQ blazar model given in Table 1. The constant-density 
BLR is confined
between 100 and 200 $R_g$,  and the Thomson depth through the BLR is $\tau_{\rm T} = 0.1$.
In Fig.\ \ref{f10}(a), the emission region of the jet is far outside the BLR, at 1000 $R_g$.
Therefore most of the BLR photons encounter
the jet from behind, and the rate of tail-on scatterings is suppressed by 
the rate factor. Consequently, the flux of the EC BLR component
is much less that the flux of the EC Disk component. 
Fig. \ref{f10}(b) presents a similar calculation, except that the BLR now 
extends to $10^3 R_g$, the same radius where the radiating jet is assumed to 
be located. 
%The optical depth through the BLR  is now increased to $\tau_{\rm T} = 0.1$. 
As expected,  the EC BLR component is significantly enhanced
compared to Fig.\ \ref{f10}(a). 

%\clearpage

\begin{figure}[h] \plotone{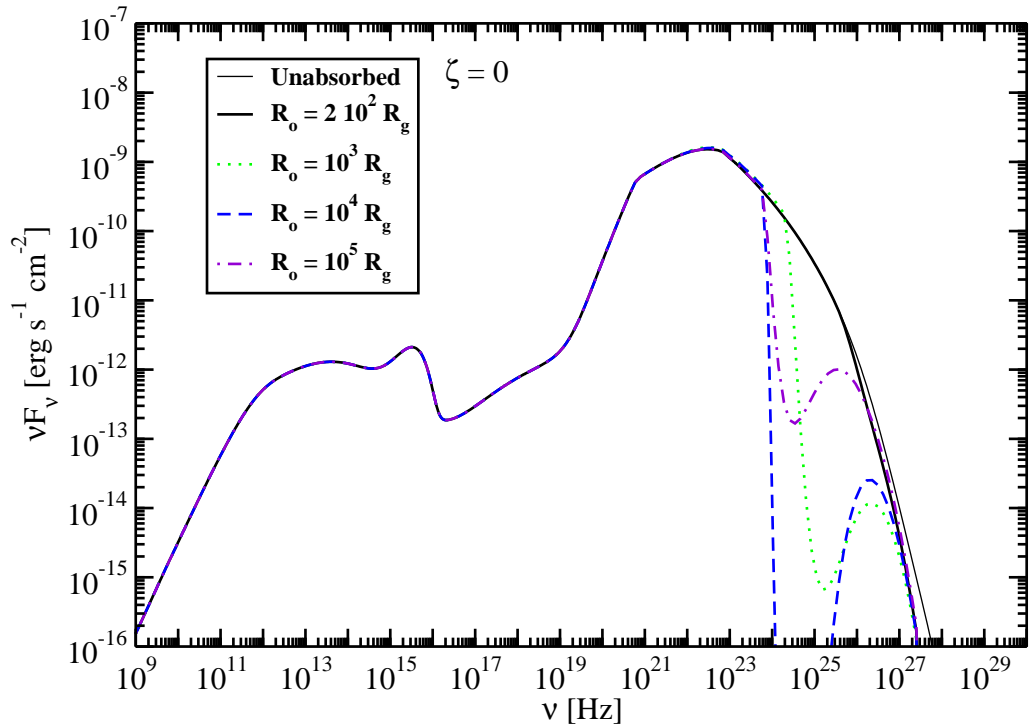}
\caption{Model blazar SED comprised of synchrotron, accretion-disk, SSC, 
EC disk, and EC BLR components, including $\g\g$ attenuation by the 
scattered BLR radiation. The jet height $r = 10^3 R_g$.
The outer radius, $R_o$, of the BLR is varied,
keeping its Thomson depth, $\tau_{\rm T} = 0.01$, constant.} 
\label{f11} 
\end{figure}

%\clearpage

There is very little $\g\g$ absorption by the accretion-disk radiation
or the BLR radiation when the jet is found outside the BLR. On the other
hand, when the jet is within the BLR, there can be significant $\g\g$ opacity.
This is shown in Fig.\ \ref{f11}, where we use the standard parameters
for the jet and a BLR with $\tau_{\rm T} = 0.01$, except that the outer
radius, $R_o$, of the BLR is varied. The effects of $\g\g$ attenuation by
the scattered radiation field are shown. When $R_o << r$, the jet height, 
then the threshold is suppressed except for the highest-energy $\gamma$ rays.
When the blob lies within the BLR, $r\lesssim R_o$,
the opacity from the scattered 
BLR radiation is significant and the $\g\g$ opacity large for $\gamma$ rays
with $\e\sim 2/\e_*$, which for our monochromatic radiation 
field with $\e_*\cong 10^{-4}$, is at $\nu \sim 5\times 10^{24}$ Hz ($\approx
40$ GeV). When $R_o\gg r$, the $\g\g$ opacity, proportional
to $r n_{sc}(\e_*;r)$, declines $\propto R_0^{-1}$ for constant Thomson 
depth, as can be seen from eq.\ (\ref{nscer}). Note that the EC $\gamma$ rays are not very sensitive to the changes
in the BLR parameters, since most of the emission is formed by the EC disk 
component.

%\clearpage

\begin{figure}[h] \plotone{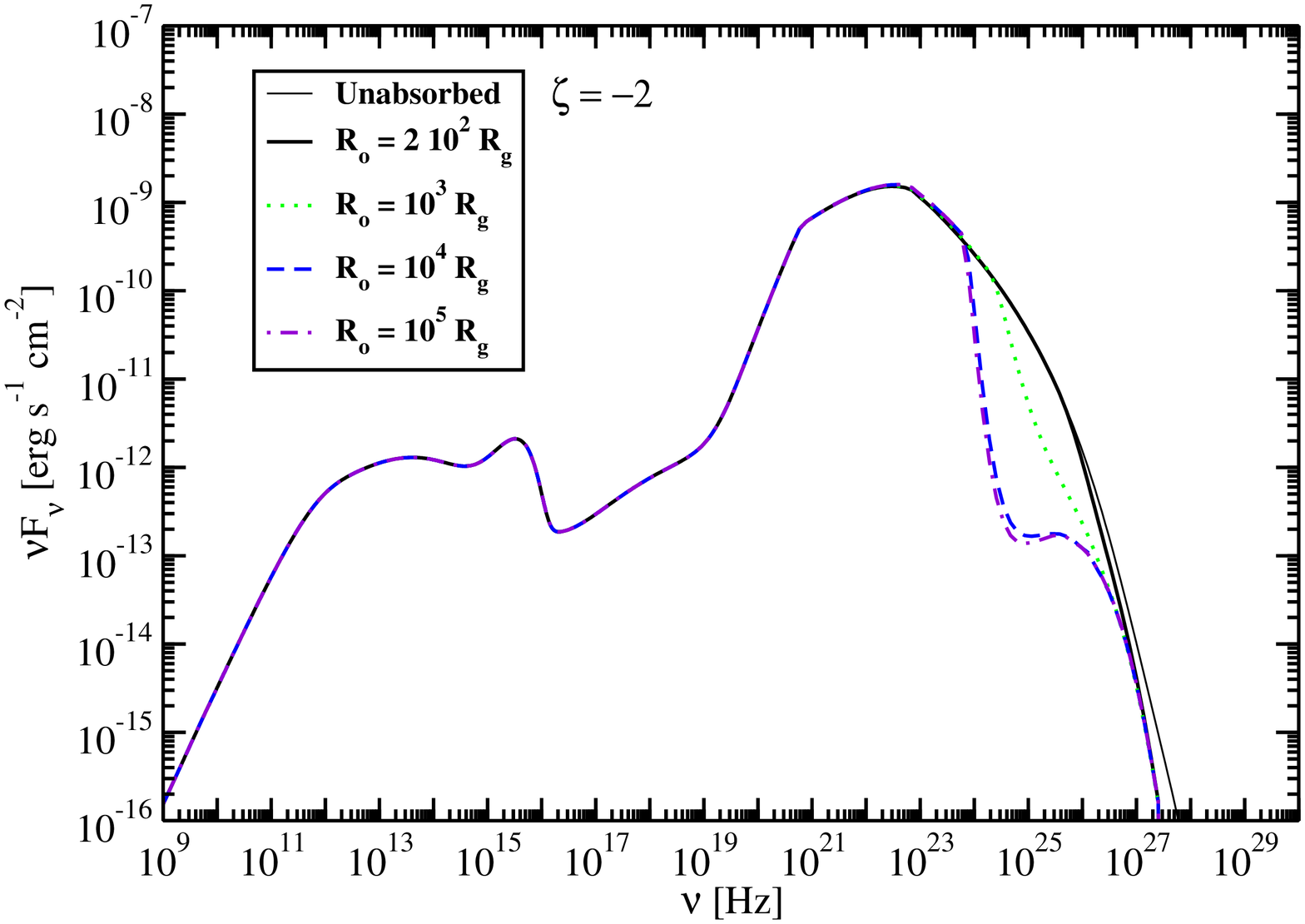} 
\caption{Same as Fig.\ \ref{f11}, except that the gradient $\zeta = -2$ 
in the density distribution.   } 
\label{f12} 
\end{figure}

%\clearpage

The effect of changing the radial gradient of BLR scattering material 
is shown in Fig.\ \ref{f12}. For a steeper density gradient, $\zeta = -2$,
 and a constant Thomson depth, the material is concentrated near the 
inner edge of the BLR at $R_i$, so the changes in the $\g\g$ opacity are 
most dramatic when $R_o \approx r$ due to geometric effects.
When $R_i \ll r \ll R_o$, the radiation field is essentially unchanged
for different values of $R_o$, and so also is the $\g\g$ opacity.

\section{Discussion and Summary}

We have presented accurate expressions for modeling synchrotron and
Compton-scattered radiation from the jets of AGN that include
target radiation fields from the accretion disk and BLR. This extends our
technique for modeling synchrotron and SSC emission \citep{fdb08} to
include external Compton scattering in a relativistic jet of thermal
radiation from the accretion disk, and accretion-disk
radiation Thomson-scattered by electrons in the BLR.  These formulae use the full
Compton cross section and are accurate throughout the Thomson and
Klein-Nishina regimes at any angle with respect to the jet axis, so 
can also be used to model $\gamma$-ray emission from radio galaxies, e.g., 
M87 \citep{aha06}. We also derive expressions for the $\g\g$ opacity
through the same scattered radiation field that serves as a target
photon source for the jet electrons. 
The expressions, eqs.\ (\ref{tauggSStilder}) and (\ref{taugge1r}), for 
opacity from the accretion-disk and scattered radiation field assume, however, that the photon 
travels along the jet axis, though it is straightforward to derive the more
general case.

In the results presented here (Figs.\ \ref{f1} -- \ref{f6} and
\ref{f10} -- \ref{f12}) we have chosen parameters for demonstration
purposes that give an exaggerated EC component, particularly a high
$\ell_{\rm Edd}$.  Lowering the disk luminosity lowers the radiation
considerably, as seen in Fig.\ \ref{f2}.  The disk radiation field in
FSRQ blazars can be seen when the nonthermal blazar radiation is in a
low state, as in the cases of 3C 279 \citep{pia99}, 3C 454.3
\citep{rai07} and, most clearly, 3C 273
\citep[e.g.,][]{mon97}. These observations can be used to assign the
accretion-disk luminosity when modeling a specific blazar, though the
disk brightness could also vary during the flaring epoch.

The models presented here do, however, have limitations.  They do not
yet include enhancements from secondary cascade radiation initiated by
$e^\pm$ pairs formed by $\g$ rays interacting with lower energy
radiation from the disk and the BLR. For the parameters considered
here, this would not make a significant difference in the calculated
SEDs because the energy flux of the attenuated radiation is a small
fraction of the escaping flux. But even in this case, the reinjected
pairs from the attenuated radiation will be isotropized if the
reinjection occurs outside the relativistic flow, and will then make
only a small contribution to the Doppler-boosted radiation.

Spectral features result from $\gamma\gamma$ absorption by BLR radiation, 
as seen in Fig.\ \ref{f12}. By assuming hard primary $\gamma$-ray 
emission components,
\citet{aha08} argue that  hard intrinsic spectra from blazars such 
as 1ES 1101-232 \citep{aha07-1101} could be formed 
through $\gamma\gamma$ attenuation. If the primary TeV radiation 
originates from an underlying
jetted electron distribution, then a consistent model requires that a
$\gamma$-ray spectrum formed by Compton-scattering processes
arises from the same radiation field responsible for $\gamma\gamma$ absorption.
%(in addition to a $\gamma$-ray spectral component formed by electrons that Compton-scatter 
%the direct accretion-disk radiation).
As our calculations show, soft Compton-scattered TeV spectra are formed due
to Klein-Nishina effects in scattering, whether from the accretion disk 
or from photons scattered by the BLR. Thus either an extremely bright
GeV component would be expected in the scenario of \citet{aha08} (cf.\ Figs.\ \ref{f11}
and \ref{f12}), 
which would easily be detected with {\em Fermi},
or a hard primary $\gamma$-ray spectrum must originate from other processes. 
Moreover, if this explanation was correct, then blazars 
with stronger broad emission lines should have harder VHE $\gamma$-ray
spectra than weak-lined BL Lacs. 

Cascade radiation induced by ultra-relativistic hadrons could inject
high-energy leptons to form a hard radiation component, though a
sufficiently dense target field for efficient photohadronic losses
will itself severely attenuate the TeV radiation
\citep{ad03}. Depending on the underlying acceleration model, a
distinctive hadronic signature could appear at $\gamma$-ray energies
only, though one would expect that both electrons and protons would be
accelerated synchronously.

Our calculations also do not as yet include absorption by the diffuse
extragalactic background light (EBL), which would be important for
EGRET $\gamma$-ray FSRQs, which have a broad redshift distribution
with a mean value $\langle z\rangle \sim 1$, larger than the mean
value $\langle z\rangle \sim 0.3$ for BL Lacs \citep{muk97}.  In cases
where the Compton-scattered spectra decline steeply due to
Klein-Nishina effects (e.g., Fig.\ \ref{f1}), the issue of EBL
absorption is secondary.  In the one FSRQ, \object{3C 279} at
$z=0.536$, detected from 80 to $> 300$ GeV energies with MAGIC
\citep{mag08}, EBL effects cannot be neglected.  For the models
studied here, the soft calculated $\gamma$-ray spectra cannot in any
case account for the measured, let alone intrinsic, spectrum of 3C
279.  Based on the simultaneous optical -- X-ray -- VHE $\gamma$-ray
SED of 3C279 during the MAGIC detection, B\"ottcher, Marscher \&
Reimer (2008, in preparation) show that one-zone leptonic models have
severe problems explaining the flare, and require either extremely
high Doppler factors or magnetic fields well below equipartition.
Correlated X-ray, {\em Fermi}, and TeV campaigns will offer
opportunities to apply the results developed in this paper.

In conclusion, we have derived expressions to model FSRQ blazars that self-consistently
include $\g\g$ attenuation on the same target photons that are Compton-scattered by
the relativistic electrons.  By assuming that the lower energy, radio -- UV emission 
is nonthermal synchrotron radiation, then the underlying electron distribution can 
be determined given the Doppler factor, magnetic field, and size scale of the 
emission region. The equations derived here can be used to calculate the 
$\gamma$-ray emission spectrum from SSC and external Compton scattering processes 
from this electron distribution. Separately, one can determine whether the 
inferred  electron distribution can be derived from specific acceleration scenarios.
Complementary to the technique of injecting electron spectra and cooling, this
method can be applied to multiwavelength data sets to analyze high-energy
processes in the jets of AGN.

%----------------------------------------------------------------------------------------
%----------------------------------------------------------------------------------------
\acknowledgements

We thank the referee for asking us to address the potential importance
of higher-order SSC fluxes, and for other helpful comments.  The work
of J.D.F. is supported by NASA {\em Swift} Guest Investigator Grant
DPR-NNG05ED411 and NASA {\em GLAST} Science Investigation
DPR-S-1563-Y, which also supported summer research by H.K. at NRL, and
a visit by M.B. to NRL. C.D.D. is supported by the Office of Naval
Research.

%!****************************************************

\end{document}